\documentclass[12pt]{amsart}
\usepackage{epsfig}
\begin{document}
\topmargin=5mm
 \oddsidemargin=6mm
 \evensidemargin=3.9mm
\baselineskip=18pt
\parindent 20pt
\begin{center}
{ \large\bf   Bilinear approach  to the
quasi-periodic wave solutions of  supersymmetric equations in superspace $\mathbb{R}_\Lambda^{2,1}$ }\\[12pt]
{\large  Engui Fan \footnote{\ \  E-mail
address: \  faneg@fudan.edu.cn}}\\[8pt]
{\small\it  School of Mathematical Sciences and Key Laboratory of
Mathematics for Nonlinear Science,\\ Fudan University, Shanghai,
200433, P.R.  China}
\end{center}
\vspace{6mm}
\rule{\textwidth}{0.5pt}\\
\baselineskip=22pt { \small {\bf Abstract}

 We devise  a  lucid and straightforward way  for explicitly constructing quasi-periodic
wave  solutions (also called multi-periodic wave solutions)  of
supersymmetric equations in superspace $\mathbb{R}_\Lambda^{2,1}$
over two-dimensional Grassmann algebra $G_1(\sigma)$. Once a
nonlinear equation is written in a bilinear form, its quasi-periodic
wave solutions can be directly obtained by using a formula.
Moreover, properties of these solutions are investigated in detail
by analyzing  their structures, plots and asymptotic behaviors. The
relations between the quasi-periodic wave solutions and soliton
solutions are rigorously established.  It is shown that the soliton
solutions can be obtained only as limiting cases of the
quasi-periodic wave solutions  under small amplitude limits in
superspace $\mathbb{R}_\Lambda^{2,1}$. We find that, in contrast to
the purely bosonic case, there is an interesting influencing band
occurred among the quasi-periodic waves under the presence of the
Grassmann variable. The quasi-periodic waves are symmetric about the
band but collapse along with the band. Furthermore, the amplitudes
of the quasi-periodic waves increase as the waves move away from the
band. The efficiency of our proposed method can be demonstrated on a
class variety of supersymmetric equations such as those considered
in this paper, $\mathcal{N}=1$ supersymmetric KdV,
Sawada-Kotera-Ramani and Ito's equations, as well as
$\mathcal{N}=2$ supersymmetric KdV  equation. } \\[8pt]
\rule{\textwidth}{0.5pt}\\[12pt]
%%%%%%%%%%%%%%%%%%%%%%%%%%%%%%%%%%%%%%%%%%%%%%%%%%%%%%%%%%%%%%%%%%%%%%%%%%%%%%%%%%%%%%%%%%%%%%%%%%%%
{\bf\large 1. Introduction}\\

The algebro-geometric solutions or finite gap solutions
 of nonlinear equations were
originally  obtained on the  KdV equation based on inverse spectral
theory and algebro-geometric method developed by pioneers such as
Novikov, Dubrovin, Mckean, Lax, Its and Matveev et al.
\cite{Nov}-\cite{Mc} in the late 1970s.  In fact, such a solution is
an expression written in terms of the Riemann theta functions. Hence
it is also called a quasi-periodic solution due to the
quasi-periodicity of the theta functions. By now this theory has
been extended to a large class of nonlinear integrable equations
including sine-Gordon equation, Camassa-Holm equation, Thirring
model equation, Kadomtsev-Petviashvili equation, Ablowitz-Ladik
lattice and Toda lattice \cite{Bel}-\cite{Fan5}.

 The quasi-periodic solutions have important applications in physics. For instance,
 they can describe the nonlinear interaction of
several modes. All the main physical characteristics of the
quasi-periodic solutions (wave numbers, phase velocities, amplitudes
of the interacting modes) are defined by a compact Riemann surface.
There are numerous applications of the finite-gap integration theory
in condensed matter physics,  state physics and fluid mechanics. For
example, in peierls state, phonon produce a finite-gap potential for
electrons, and  the peierls state is a lattice of solutions at low
densities of electrons \cite{Bel}.  A most famous mechanical system,
the Kowalewski top,  was the focus of interest in the 19th century.
The equation of motion of the top can be solved through finite-gap
theory  \cite{Bel}.  A problem of fundamental interest in fluid
mechanics is to provide an accurate description of waves on a water
surface. The Kadomtsev-Petviashvili  equation is known to describe
the evolution of waves in shallow water and admits a large family of
quasi-periodic solutions. Each
 solution has $N$ independent phases.  Experiments demonstrate the existence of
 genuinely two-dimensional shallow water waves that are full periodic in two spatial directions and time.
  The comparisons with experiments showed that the two-periodic wave solutions of
the KP equation  describe shallow water waves with much accuracy
\cite{HA1,  HA2}.

 The algebro-geometric theory, however, needs Lax pairs and involves complicated calculus on the Riemann surfaces.
 It is  rather difficult to directly determine
the characteristic parameters of waves such as frequencies and phase
shifts for a function of given wave-numbers and amplitudes. On the
other hand, the bilinear derivative method developed by Hirota is a
powerful approach for constructing exact solution of nonlinear
equations.  Once a nonlinear equation is written in bilinear forms
by a dependent variable transformation, then multi-soliton solutions
are usually obtained \cite{Hirota1}--\cite{Sa}. It was based on
Hirota forms that Nakamura proposed a convenient way to construct a
kind of quasi-periodic  solutions of  nonlinear equations \cite{Na1,
Na2, Hirota0},   where  the  periodic  wave solutions of the KdV
equation and the Boussinesq  equation were obtained. Such a method
indeed exhibits some advantages over algebro-gometric methods. For
example, it does not need any Lax pairs and Riemann surface for the
considered equation, allows the explicit construction of
multi-periodic wave solutions, only relies on the existence of the
Hirota's bilinear form, as well as all parameters appearing in
Riemann matrix are arbitrary. Recently, further development was made
to investigate the discrete Toda lattice, (2+1)-dimensional
Kadomtsev-Petviashvili equation  and Bogoyavlenskii's   breaking
soliton equation \cite{Dai}-\cite{Ma2}. Indeed there are some
differences between quasi-periodic solutions and algebro-geometric
solutions. A quasi-periodic solution needs not be an
algebro-geometric one. Sometimes a quasi-periodic solution may not
correspond to any Riemann surface and  is generically associated
with infinite bands, not just finitely-many, for instance  with a
Riemann surface of infinite genus.

The concept of supersymmetry  was originally introduced and
developed for applications in elementary particle physics thirty
years ago \cite{Ramond}--\cite{Wess}. It is found that supersymmetry
can be  applied to a variety of problems such as relativistic,
non-relativistic physics and nuclear physics. In recent years,
supersymmetry has been a subject of considerable interest both in
physics and mathematics.  The mathematical formulation of the
supersymmetry is based on the introduction of Grassmann variables
along with the standard ones \cite{Berezin}. In a such way, a number
of well known mathematical physical equations have been generalized
into the supersymmetric analogues, such as supersymmetric versions
of sine-Gordon, KdV, KP hierarchy, Boussinesq, MKdV  etc.
\cite{Manin}--\cite{Ghosh}.  It has been shown that these
supersymmetric integrable systems possess bi-Hamiltonian structure,
Painleve property, infinite many symmetries, Darboux transformation,
Backlund transformation, bilinear form and multi-soliton solutions.
The systematic bilinear transcription of supersymmetric equations
was introduced by Carstea \cite{Carstea, Carstea1}. This required an
extension of the Hirota's bilinear operator to supersymmetric case.
Despite this bilinearization of  supersymmetric equations, the
standard construction did not lead to malti-soliton solutions. In
recent years, Carsta,  Liu, Ghosh et al. have done much on the
construction of soliton solutions of supersymmetric equations
\cite{Carstea}--\cite{Ghosh} . However,  the quasi-periodic
solutions of the supersymmetric systems, which can be considered as
a generalization of the soliton solutions,  are still not available
(both by algebro-geometric method and by bilinear methods or others)
to the knowledge of the author.

  The motivation of this paper is to show how the quasi-periodic wave solutions of nonlinear
   supersymmetric equations can be constructed with Hirota's bilinear method in superspace.
To achieve this aim,  we devise a Riemannn theta function formula,
which actually provides us a lucid and straightforward  way for
applying in a class of nonlinear supersymmetric equations. Once a
nonlinear equation is written in bilinear forms, then the
quasi-periodic wave solutions of the nonlinear equation can be
obtained directly  by using the formula.   This method considerably
improves the key steps of the existing
  methods, where repetitive recursion and computation must be preformed for each
equation \cite{Dai}-\cite{Ma2}.   As illustrative example, we shall
  construct quasi-periodic wave   solutions to the $\mathcal{N}=1$ supersymmetric Sawada-Kotera-Ramani equation
  and $\mathcal{N}=2$ supersymmetric KdV equation.

  The organization of this paper is as follows. In section 2, we briefly
  give  some properties on superspace and super-Hirota bilinear  operators.
  In  section 3, we introduce a  super Riemann theta function and
  discuss its quasi-periodicity. In particular,  we  provide a key formula for constructing periodic wave solutions
 of supersymmetric equations.  As applications of our
method, in section 4 and section 5,   we construct one- and
two-periodic wave solutions to the $\mathcal{N}=1$ supersymmetric
Sawada-Kotera-Ramani equation   and $\mathcal{N}=2$ supersymmetric
KdV equation, respectively. The propagation of the quasi-periodic
waves are  displayed with help of software Mathematica. In addition,
we further present a simple and effective limiting procedure to
analyze asymptotic behavior of the periodic wave solutions.  It is
rigorously shown that the quasi-periodic wave solutions tend to the
soliton solutions under small amplitude limits. At last, we briefly
discuss the conditions on the construction of multi-periodic wave
solutions of supersymmetric equations in section 6.\\[12pt]
%%%%%%%%%%%%%%%%%%%%%%%%%%%%%%%%%%%%%%%%%%%%%%%%%%%%%%%%%%%%%%%%%%%%%%%%
%%%%%%%%%%%%%%%%%%%%%%%%%%%%%%%%%%%%%%%%%%%%%%%%%%%%%%%%%%%%%%%%%%%%%%%%%%%%%%
{\bf\large  2.  Super space  and super-Hirota bilinear form }\\

To fix the notations  and make our presentation self-contained,  we
briefly recall  some properties about superanalysis  and
super-Hirota bilinear operators.  The details about superanalysis
refer, for instance, to Vladimirov's work \cite{Vlad1, Vlad2}.

A linear space $\Lambda$ is called $Z_2$-graded if it represented as
a direct sum of two subspaces
$$\Lambda=\Lambda_0\oplus\Lambda,$$
where elements of the spaces $\Lambda_0$ and $\Lambda_1$ are
homogeneous. We assume that  $\Lambda_0$ is a subspace consisting of
even elements  and  $\Lambda_1$ is a subspace consisting of odd
elements. For the element $f\in\Lambda$ we denote by $f_0$ and $f_1$
its even and odd components. A parity function is introduced on the
$\Lambda$, namely,
$$|f|=\left\{\begin{matrix}0, \ \ {\rm if} \ \ f\in \Lambda_0,
\\ \ 1, \ \ {\rm if }\ \ f\in \Lambda_1.\end{matrix}\right.
$$

We introduce an annihilator of the set of odd elements by setting
$$^{\perp}\Lambda_1=\{\lambda\in\Lambda: \lambda\Lambda_1=0\}.$$

A superalgebra  is a $Z_2$-graded space
$\Lambda=\Lambda_0\oplus\Lambda$ in which, besides usual operations
of addition and multiplication by numbers, a product of elements is
defined with the usual distribution law:
$$a(\alpha b+\beta c)=\alpha ab+\beta ac, \ \ (\alpha b+\beta
c)a=\alpha ba+\beta ca,$$ where $a, b, c\in\Lambda$ and $\alpha,
\beta\in \mathbb{C}.$  Moreover,  a structure on $\Lambda$ is
introduced of an associative algebra with a unite $e$ and even
multiplication i.e., the product of two even and two odd elements is
an even element and the product of an even element by an odd one is
an odd element: $|ab|=|a|+|b|$ mod (2).

 A commutative superalgebra with unit $e=1$ is
called a finite-dimensional  Grassmann algebra if it contains a
system of anticommuting generators $\sigma_j, j=1, \cdots, n$ with
the property: $\sigma_j\sigma_k+\sigma_k\sigma_j=0, \ j,
k=1,2,\cdots, n$, in particular, $\sigma_j^2=0$. The Grassmann
algebra will be denote by $G_n=G_n(\sigma_1,\cdots, \sigma_n)$.

The monomials $\{e_0, e_i=\sigma_{j_1}\cdots\sigma_{j_n}\}$,
$j=(j_1<\cdots<j_n)$ form a basis in the Grassmann algebra $G_n$,
$\dim G_n=2^n$.  Then it follows that any element of $G_n$ is a
linear combination of monomials $\sigma_{j_1}\cdots\sigma_{j_k}, \
j_1<\cdots<j_k$, that is,
$$f=f_0+\sum_{k\geq 0}\sum_{j_1<\cdots<j_k}f_{j_1\cdots j_k}\sigma_{j_1}\cdots\sigma_{j_k},$$
where the coefficients $f_{j_1\cdots j_k}\in \mathbb{C}$.

{\bf Definition 1.}  Let $\Lambda=\Lambda_0\oplus\Lambda$ be a
commutative Banach superalgebra, then the Banach space
$$\mathbb{R}_{\Lambda}^{m,n}=\Lambda_0^m\times\Lambda_1^n$$
is called a superspace of dimension $(m,n)$ over $\Lambda$.  In
particular, if $\Lambda_0=\mathbb{C}$ and $\Lambda_1=0$, then
$\mathbb{R}_{\Lambda}^{m,n}=\mathbb{C}^m.$

 A function $f(\boldsymbol{x}):
\mathbb{R}_{\Lambda}^{m,n}\rightarrow \Lambda$ is said to be
superdifferentiable at the point $x\in \mathbb{R}_{\Lambda}^{m,n}$,
if there exist elements $F_j(\boldsymbol{x})$ in $\Lambda, \ j=1,
\cdots, m+n$, such that
$$f(\boldsymbol{x}+\boldsymbol{h})
=f(\boldsymbol{x})+\sum_{j=1}^{m+n}\langle
F_j(\boldsymbol{x}),h_j\rangle+o(\boldsymbol{x},\boldsymbol{h}),$$
where $\boldsymbol{x}=(x_1, \cdots, x_m, x_{m+1}, \cdots, x_{n})$
with components $ x_j, j=1,\cdots, m$ being even variable and
$x_{m+j}=\theta_j, j=1,\cdots, n$  being Grassmann odd ones. The
vector $\boldsymbol{h}=(h_1, \cdots, h_m$, $h_{m+1}, \cdots,
h_{m+n})$ with $ (h_1, \cdots, h_m)\in\Lambda_0^m$ and $(h_{m+1},
\cdots, h_{m+n})\in\Lambda_1^n$. Moreover,
$$\lim_{\parallel \boldsymbol{h}\parallel\rightarrow 0}\frac{\parallel o(\boldsymbol{x},\boldsymbol{h})
\parallel}{\parallel \boldsymbol{h}\parallel}
\longrightarrow 0.$$ The $F_j(\boldsymbol{x})$ are called the super
partial derivative of $f$ with respect to $x_j$ at the point
$\boldsymbol{x}$ and are denoted, respectively, by
$$\frac{\partial f(\boldsymbol{x})}{\partial x_j}=F_j(\boldsymbol{x}),\ j=1,\cdots, m+n.$$
The derivatives $\frac{\partial f(\boldsymbol{x})}{\partial x_j}$
with respect to even variables  $x_j, \ j=1,2,\cdots n$ are uniquely
defined. While the derivatives $\frac{\partial
f(\boldsymbol{x})}{\partial \theta_j}$ to odd variables
$\theta_j=x_{j+n}, \ j=1,2,\cdots m$ are not uniquely defined, but
with an accuracy to within an addition constant
$c\sigma_1\cdots\sigma_n, c\in \mathbb{C}$ from an annihilator
$^\perp G_n$ of finite-dimensional Grassmann algebra $G_n$.

The super derivative also satisfies Leibniz formula
$$\frac{\partial (f(\boldsymbol{x})g(\boldsymbol{x}))}{\partial x_j}=\frac{\partial f(\boldsymbol{x})}
{\partial x_j}g(\boldsymbol{x})
+(-1)^{|x_j||f|}f(\boldsymbol{x})\frac{\partial
g(\boldsymbol{x})}{\partial x_j}, \ j=1, \cdots, m+n.\eqno(2.1)$$

 Denote by  $\mathcal{P}(\Lambda_1^n,
\Lambda)$ the set of polynomials defined on $\Lambda_1^n$ with value
in $\Lambda$. We say that  a super integral is a map $I:
\mathcal{P}(\Lambda_1^n, \Lambda)\rightarrow \Lambda$ satisfying the
following condition is an super integral about Grassmann variable

(1) A linearity: $I(\mu f+\nu g)=\mu I(f)+\nu I(g), \ \mu,
\nu\in\Lambda, \ f, g\in\mathcal{P}(\Lambda_1^n, \Lambda); $

(2) translation invariance: $I(f_{\xi})=I(f)$, where
$f_{\xi}=f(\boldsymbol{\theta}+\boldsymbol{\xi})$ for all
$\boldsymbol{\xi}\in\Lambda_1^n$, $f\in\mathcal{P}(\Lambda_1^n,
\Lambda).$

We denote $I(\theta^\varepsilon)=I_\varepsilon$, where $\varepsilon$
belongs to the set of multiindices
$N_n=\{\boldsymbol{\epsilon}=(\varepsilon_1, \cdots, \varepsilon_n),
\varepsilon_j=0,1,
\boldsymbol{\theta}^\varepsilon=\theta_1^{\varepsilon_1}\cdots\theta_n^{\varepsilon_n}\not\equiv
0\}$. In the case when $I_\varepsilon=0, \varepsilon\in N_n,
|\varepsilon|\leq n=n-1$, such kind of integral has the form
$$I(f)=J(f)I(1,\cdots,1),$$
where
$$J(f)=\frac{\partial^nf(0)}{\partial\theta_1\cdots\partial\theta_n}.$$
Since the derivative is defined with an accurcy to with an additive
constant form the annihilator $^\perp L_n$,
$L_n=\{\theta_1\cdots\theta_n, \boldsymbol{\theta}\in
\Lambda_1^n\}$, it follows that $J:\mathcal{P}\rightarrow
\Lambda/^\perp L_n$ is single-valued mapping. This mapping also
satisfies the conditions 1 and 2, and therefore we shall call it an
integral and denote
$$J(f)=\int f(\boldsymbol{\theta})d\boldsymbol{\theta}=\int\theta_1\cdots\theta_n d\theta_1\cdots d\theta_n,$$
which has properties:
$$\begin{aligned}
&\int\theta_1\cdots\theta_n d\theta_1\cdots d\theta_n=1,\\
&\int\frac{\partial f}{\partial\theta_j}d\theta_1\cdots
d\theta_n=0,\ j=1, \cdots, n.\\
&\int f(\boldsymbol{\theta})\frac{\partial
g(\boldsymbol{\theta})}{\partial\theta_j}d\boldsymbol{\theta}=
(-1)^{1+|g|}\int \frac{\partial
f(\boldsymbol{\theta})}{\partial\theta_j}g(\boldsymbol{\theta})d\boldsymbol{\theta}.
\end{aligned}\eqno(2.2)$$

In this paper, we consider functions with two ordinary even
variables $x, t$ and a Grassmann odd variable $\theta$. The
associated  space
$\mathbb{R}_{\Lambda}^{2,1}=\Lambda_0^2\times\Lambda_1$ (we may take
$\Lambda_0=\mathbb{R}$ or $\mathbb{C}$) is a superspace over
Grassmann algebra $G_1(\sigma)=G_{1,0}\oplus G_{1,1}$, whose
elements have the form
 $$f=f_0+f_1\sigma.$$
where $e=1$ is a unit, $\sigma$ is anticommuting generator.
 The monomials $\{1, \sigma\}$ form a basis of the
$G_1(\sigma)$, dim$G_1(\sigma)=2$. Therefore, any $\mu\in G_{1,1}$
have the form $\mu=\beta\sigma, \ \beta\in \mathbb{C}$.  Under
traveling wave frame in space $\mathbb{R}_{\Lambda}^{2,1}$, the
 phase variable  should have  the form
$$\xi=\alpha x+\omega t+\beta\theta\sigma. $$

For the functions $f(x,t,\theta), g(x,t,\theta):
\mathbb{R}_{\Lambda}^{2,1}\rightarrow \Lambda$, the  Hirota bilinear
differential operators $D_x$ and $D_t$ about ordinary variables $x,
t$ are defined by
\begin{eqnarray*}
&&D_x^mD_t^n f(x,t,\theta)\cdot g(x, t,
\theta)=(\partial_x-\partial_{x'})^m(\partial_t-\partial_{t'})^n
f(x, t, \theta) g(x', t', \theta)|_{x'=x, t'=t}.
\end{eqnarray*}
 The super-Hirota bilinear operator is defined as
\cite{Carstea}
$$ S_x^N f(x,t,\theta)\cdot
g(x,t,\theta)=\sum_{j=0}^N(-1)^{j|f|+\frac{1}{2}j(j+1)}\left[\begin{matrix} N\\
j\end{matrix}\right]\mathfrak{D}^{N-j}f(x,t,\theta)\mathfrak{D}^jg(x,t,\theta),$$
 where  the differential operator $\mathfrak{D}=\partial_{\theta}+\theta\partial_x$
  is the super derivative, and the super
binomial coefficients are defined by
$$\left[\begin{matrix} N\\
j\end{matrix}\right]=\left\{\begin{matrix}\left(\begin{matrix}
[N/2]\cr [j/2]\end{matrix}\right), {\rm if} \ \ (N, j)\not=(0,1)\ \
{\rm mod}\ \ 2,\\ \\ 0, \ \  \ \ \  \ \ {\rm otherwise}\ \ \ \ \ \ \
\ \ \ \ \ \ \ .\end{matrix}\right.$$ $[k]$ is the integer part of
the real number $k$ ($[k]\leq k\leq [k]+1$).

We point out here that throughout this paper the natural number $N$
(which will denote  powers, the number of phase variables,  number
of terms etc.) is different form $\mathcal{N}$ which is related to
supersymmetry or superspace.

 {\bf Proposition 1.}  Suppose that functions $f(x,t,\theta), g(x,t,\theta):
\mathbb{R}_{\Lambda}^{2,1}\rightarrow \Lambda$, then  Hirota
bilinear operators $D_x, D_t$ and super-Hirota bilinear operator
$S_x$ have properties \cite{Carstea}
\begin{eqnarray*}
&&S_x^{2N}f\cdot
  g =D_x^N f\cdot g,\\
&&D_x^mD_t^n   e^{\xi_1}\cdot
  e^{\xi_2}=(\alpha_1-\alpha_2)^m(\omega_1-\omega_2)^n
  e^{\xi_1+\xi_2},\\
  &&S_x e^{\xi_1}\cdot
  e^{\xi_2}=[\sigma(\beta_1-\beta_2)+\theta(\alpha_1-\alpha_2)]
  e^{\xi_1+\xi_2},
\end{eqnarray*}
where  $\xi_j=\alpha_jx+\omega_jt+\beta_j\theta\sigma+\delta_j$, $\
 \alpha_j,  \omega_j,  \sigma_j, \delta_j\in\Lambda_0$ are
parameters, $ j=1,2$. In fact, the third formula above is defined
with an accuracy to within an addition constant of the $c\sigma\in
^\perp\Lambda_1$. More generally, we have
$$\begin{aligned}
&F(S_x, D_x, D_t)e^{\xi_1}\cdot
  e^{\xi_2}  =F(\sigma(\beta_1-\beta_2)+\theta(\alpha_1-\alpha_2),\alpha_1-\alpha_2,\omega_1-\omega_2)
  e^{\xi_1+\xi_2},\end{aligned}\eqno(2.3)$$
where $F(S_t, D_x, D_t)$ is a polynomial about operators $S_t, D_x$
and $ D_t$. This properties  are useful in deriving Hirota's
bilinear form and constructing   the quasi-periodic wave solutions
of the supersymmetric equations.\\[12pt]
%%%%%%%%%%%%%%%%%%%%%%%%%%%%%%%%%%%%%%%%%%%%%%%%%%%%%%%%%%%%%%%%%%%%%%%%%%%%%%
{\bf\large  3.  Super  Riemann theta
function and  addition formulae}\\

In the following,  we introduce a  multi-dimensional super Riemann
theta function on superspace $\mathbb{R}_{\Lambda}^{2,1}$ and
discuss its quasi-periodicity, which plays a central role in the
construction of quasi-periodic solutions of supersymmetric
equations. The multi-dimensional Riemann theta function reads
$$
\vartheta(\boldsymbol{\xi},
\boldsymbol{\varepsilon},\boldsymbol{s}|\boldsymbol{\tau})=\sum_{\boldsymbol{n}\in
\mathbb{{Z}}^N}\exp\{2\pi
i\langle\boldsymbol{\xi}+\boldsymbol{\varepsilon},\boldsymbol{n}+\boldsymbol{s}\rangle-\pi
\langle\boldsymbol{\tau}
(\boldsymbol{n}+\boldsymbol{s}),\boldsymbol{n}+\boldsymbol{s}\rangle\}.\eqno(3.1)$$
Here the integer value vector $\boldsymbol{n}=(n_1,\cdots,n_N)^T\in
\mathbb{Z}^N$, complex parameter vectors
$\boldsymbol{s}=(s_1,\cdots,s_N)^T,
\boldsymbol{\varepsilon}=(\varepsilon_1,\cdots,\varepsilon_N)^T\in
\mathbb{{C}}^N$. The complex phase variables $\boldsymbol{\xi}
=(\xi_1, \cdots, \xi_N)^T, \
\xi_j=\alpha_jx+\omega_jt+\beta_j\theta\sigma+\delta_j$, $\
\alpha_j, \omega_j,\beta_j, \delta_j\in\Lambda_0$, $  j=1, 2,
\cdots, N$, where $x, t$ are ordinary variables and $\theta$ is
Grassmann variable. Moreover, for two vectors $\boldsymbol{f}=(f_1,
\cdots, f_N)^T$ and $\boldsymbol{g}=(g_1, \cdots, g_N)^T$,
 their  inner product is defined by
 $$\langle \boldsymbol{f}, \boldsymbol{g}\rangle=f_1g_1+f_2g_2+\cdots+f_Ng_N.$$
 The $\boldsymbol{\tau}=(\tau_{ij})$ is  a positive
definite and real-valued symmetric $N\times N$ matrix, which is
independent of $\theta$ and $\sigma$ in superspace
$\mathbb{R}_\Lambda^{2,1}$. The entries $\tau_{ij}$ of the period
matrix $\boldsymbol{\tau}$ can be considered as free parameters of
the theta function (3.1).

   In
this paper, we take the $\tau$ to be pure imaginary matrix to make
the theta function (3.1) real-valued.   In the definition of the
theta function (3.1), for  the case
$\boldsymbol{s}=\boldsymbol{\varepsilon}=\boldsymbol{0}$, hereafter
we use
$\vartheta(\boldsymbol{\xi},{\boldsymbol{\tau}})=\vartheta(\boldsymbol{\xi},\boldsymbol{0},
\boldsymbol{0}|\boldsymbol{\tau})$ for simplicity.  Moreover, we
have  $\vartheta(\boldsymbol{\xi},\boldsymbol{\varepsilon},
\boldsymbol{0}|\boldsymbol{\tau})=\vartheta(\boldsymbol{\xi}+\boldsymbol{\varepsilon},
\boldsymbol{\tau})$.  It is obvious that the Riemann theta function
(3.1)  converges absolutely and superdifferentiable on superspace
$\mathbb{R}_{\Lambda}^{2,1}$.

{\bf Remark 1.}  The period matrix $\boldsymbol{\tau}$ here is
different form algebro-geometric theory discussed in
\cite{Nov}-\cite{Fan2}, where it is usually constructed via  a
compact Riemann surface $\Gamma$ of genus $N\in\mathbb{N}$. We take
two sets of regular cycle paths: $a_1, a_2, \cdots, a_{N}$; $b_1,
b_2, \cdots, b_{N}$ on $\Gamma$ in such a way that  the intersection
numbers of cycles satisfies
$$a_k\circ a_j=b_k\circ b_j=0, a_k\circ b_j=\delta_{kj}, \ \ k, j=1,\cdots, N.$$
We choose the  normalized holomorphic  differentials $\omega_j, j=1,
\cdots, N$ on $\Gamma$  and let
$$a_{jk}=\int_{a_k}{{\omega}}_j, \ \ b_{jk}=\int_{b_k}{{\omega}}_j,$$
then  $N\times N$ matrices $\boldsymbol{A}=(a_{jk})$ and
$\boldsymbol{B}=(b_{jk})$ are invertible.  Define matrices
$\boldsymbol{C}$ and $\boldsymbol{\tau}$ by
$$\boldsymbol{C}=(c_{jk})=\boldsymbol{A}^{-1},\ \  \boldsymbol{\tau}=(\tau_{jk})=\boldsymbol{A}^{-1}\boldsymbol{B}.$$
It is can be shown that the matrix $\boldsymbol{\tau}$ is symmetric
and has positive definite imaginary part. However, we see that the
entries in such a matrix $\boldsymbol{\tau}$ are not free and
difficult to be explicitly given. $\square$

{\bf Definition 2.}  A function  $g(\boldsymbol{x},t)$ on
$\mathbb{C}^N\times\mathbb{C}$  is said to be quasi-periodic in $t$
with fundamental periods  $T_1, \cdots, T_k\in \mathbb{C}$ if
 $T_1, \cdots, T_k$ are linearly dependent over $\mathbb{Z}$ and there exists a function
  $G(\boldsymbol{x},t)\in \mathbb{C}^N\times\mathbb{C}^k$ such that
$$ G(\boldsymbol{x}, y_1,\cdots, y_j+T_j, \cdots, y_k)=G(\boldsymbol{x},y_1,\cdots, y_j, \cdots, y_k ),
 \ \ {\rm for\ all}\ y_j\in \mathbb{C}, \ j=1, \cdots, k.$$
$$ G(\boldsymbol{x}, t,\cdots, t, \cdots, t )=g(x, t). $$
In particular, $g(\boldsymbol{x},t)$ becomes periodic with $T$ if
and only if $T_j=m_jT$. $\square$

 Let's first see the periodicity of the theta function $\vartheta(\boldsymbol{\xi}, \boldsymbol{\tau})$.

 {\bf Proposition 2.} \cite{Far} Let $\boldsymbol{e_j}$ be the $j-$th
 column of  $N\times N$  identity matrix $I_N$;
 ${\tau_j}$ be the $j-$th column of $\boldsymbol{\tau}$, and $\tau_{jj}$ the
 $(j,j)$-entry of $\boldsymbol{\tau}$. Then the theta function $\vartheta(\boldsymbol{\xi}, \boldsymbol{\tau})$ has the periodic properties
$$\begin{aligned}
&\vartheta(\boldsymbol{\xi}+\boldsymbol{e_j}+i\boldsymbol{\tau_j},\boldsymbol{\tau})=\exp(-2\pi
i\xi_j+\pi \tau_{jj})\vartheta(\boldsymbol{\xi},\boldsymbol{\tau}).
\end{aligned}\eqno(3.2)$$

  The theta function $\vartheta(\boldsymbol{\xi},\boldsymbol{\tau})$ which satisfies the condition (4.4)  is called
 a multiplicative function. We regard the vectors
$\{\boldsymbol{e_j},\ \ j=1, \cdots, N\}$ and
$\{i\boldsymbol{\tau_j}, \ \ j=1, \cdots, N\}$ as periods of the
theta function $\vartheta(\boldsymbol{\xi},\boldsymbol{\tau})$ with
multipliers $1$ and $\exp({-2\pi i\xi_j+\pi \tau_{jj}})$,
respectively. Here, only the first $N$ vectors are actually periods
of the theta function $\vartheta(\boldsymbol{\xi},
\boldsymbol{\tau})$, but  the last $N$ vectors are  the periods of
the functions  $\partial^2_{\xi_k,\xi_l}\ln
\vartheta(\boldsymbol{\xi},\boldsymbol{\tau})$ and $
\partial_{\xi_k}\ln[\vartheta(\boldsymbol{\xi}+\boldsymbol{e},
\boldsymbol{\tau})/\vartheta(\boldsymbol{\xi}+\boldsymbol{h},\boldsymbol{\tau})],
\ k, l=1, \cdots, N$.

{\bf Proposition 3.} Let $\boldsymbol{e_j}$ and
$\boldsymbol{\tau_j}$ be defined as above proposition 2. The
meromorphic functions $f(\boldsymbol{\xi})$ on
$\mathbb{R}_\Lambda^{2,1}$ are as follow
$$\begin{aligned}
&(i) \ \ \ \
 \  f(\boldsymbol{\xi})=\partial_{\xi_k\xi_l}^2\ln\vartheta(\boldsymbol{\xi},
 \boldsymbol{\tau}),\ \ \boldsymbol{\xi}\in C^N, \ \ \ k,  l=1, \cdots,
 N,
\end{aligned}$$
$$\begin{aligned}
&(ii) \ \ \ \
f(\boldsymbol{\xi})=\partial_{\xi_k}\ln\frac{\vartheta(\boldsymbol{\xi}+
\boldsymbol{e},\boldsymbol{\tau})}{
\vartheta(\boldsymbol{\xi}+\boldsymbol{h},\boldsymbol{\tau})},\ \
\boldsymbol{\xi},\ \boldsymbol{e},\ \boldsymbol{h}\in C^N, \ \ j=1,
\cdots, N.
\end{aligned}$$
   then in all two cases (i) and  (ii), it holds that
$$\begin{aligned}
&f(\boldsymbol{\xi}+\boldsymbol{e_j}+i\boldsymbol{\tau_j})=f(\boldsymbol{\xi}),
\ \ \ \boldsymbol{\xi}\in C^N,\ \ \ j=1, \cdots, N.
\end{aligned}\eqno(3.3)$$

{\it Proof.} By using (3.2),  it is easy to see that
$$\begin{aligned}
&\frac{\vartheta'_{\xi_k}(\boldsymbol{\xi}+\boldsymbol{e_j}+i\boldsymbol{\tau_j},\boldsymbol{\tau})}
{\vartheta(\boldsymbol{\xi}+\boldsymbol{e_j}+i\boldsymbol{\tau_j},\boldsymbol{\tau})}
=-2\pi
i\delta_{jk}+\frac{\vartheta'_{\xi_k}(\boldsymbol{\xi},\boldsymbol{\tau})}{\vartheta(\boldsymbol{\xi},\boldsymbol{\tau})},
\end{aligned}$$
or equivalently
$$\begin{aligned}
&\partial_{\xi_k}\ln\vartheta(\boldsymbol{\xi}+\boldsymbol{e_j}+i\boldsymbol{\tau_j},\boldsymbol{\tau})=-2\pi
i\delta_{jk}+\partial_{\xi_k}\ln
\vartheta(\boldsymbol{\xi},\boldsymbol{\tau}).
\end{aligned}\eqno(3.4)$$
Differentiating  (3.4) with respective to $\xi_l$ again immediately
proves the formula (3.3) for the case (i).  The formula (3.4) can be
proved  for the case (ii) in a similar manner.  $\square$

{\bf Theorem 1.}   Suppose that
$\vartheta(\boldsymbol{\xi},\boldsymbol{\varepsilon'},
\boldsymbol{0}|\boldsymbol{\tau})$ and $
\vartheta(\boldsymbol{\xi},\boldsymbol{\varepsilon},
\boldsymbol{0}|\boldsymbol{\tau})$ are two Riemann theta functions
on $\mathbb{R}_\Lambda^{2,1}$, in which
$\boldsymbol{\varepsilon}=(\varepsilon_1, \dots, \varepsilon_N)$,
$\boldsymbol{\varepsilon'}=(\varepsilon_1', \dots, \varepsilon_N')$,
and  $\boldsymbol{\xi}=(\xi_1, \cdots, \xi_N)$,
$\xi_j=\alpha_jx+\omega_jt+\beta_j\theta\sigma+\delta_j, \ \ j=1, 2,
\cdots, N$. Then  Hirota bilinear operators $D_x, D_t$ and
super-Hirota bilinear operator $S_x$ exhibit the following perfect
properties when they act on a pair of theta functions
$$\begin{aligned}
&D_x \vartheta(\boldsymbol{\xi},\boldsymbol{\varepsilon'},
\boldsymbol{0}|\boldsymbol{\tau})\cdot
  \vartheta(\boldsymbol{\xi},\boldsymbol{\varepsilon}, \boldsymbol{0}|\boldsymbol{\tau})\\
  &=\left[\sum_{\boldsymbol{\mu}}\partial_x\vartheta(2\boldsymbol{\xi},\boldsymbol{\varepsilon'}-\boldsymbol{\varepsilon},
  -\boldsymbol{\mu}/2|2\boldsymbol{\tau})|_{\boldsymbol{\xi}=\boldsymbol{0}}\right]
  \vartheta(2\boldsymbol{\xi},\boldsymbol{\varepsilon'}+\boldsymbol{\varepsilon},\boldsymbol{\mu}/2|2\boldsymbol{\tau}),
\end{aligned}\eqno(3.5)$$
$$\begin{aligned}
&S_x  \vartheta(\boldsymbol{\xi},\boldsymbol{\varepsilon'},
\boldsymbol{0}|\boldsymbol{\tau})\cdot
  \vartheta(\boldsymbol{\xi},\boldsymbol{\varepsilon}, \boldsymbol{0}|\boldsymbol{\tau})\\
  &=\left[\sum_{\boldsymbol{\mu}}\mathfrak{D}_x\vartheta(2\boldsymbol{\xi},\boldsymbol{\varepsilon'}-\boldsymbol{\varepsilon},
  -\boldsymbol{\mu}/2|2\boldsymbol{\tau})|_{\boldsymbol{\xi}=\boldsymbol{0}}\right]
  \vartheta(2\boldsymbol{\xi},\boldsymbol{\varepsilon'}+\boldsymbol{\varepsilon},\boldsymbol{\mu}/2|2\boldsymbol{\tau}),
\end{aligned}\eqno(3.6)$$
 where $\boldsymbol{\mu}=(\mu_1, \cdots, \mu_N)$,  and  the notation  $\sum_{\boldsymbol{\mu}}$
  represents $2^N$ different transformations corresponding to all possible
  combinations  $\mu_1=0,1;  \cdots;  \mu_N=0,1$.

  In general, for a polynomial operator $F(S_x, D_x,
D_t)$ with respect to  $S_x, D_x$ and $ D_t$, we have the following
useful formula
$$\begin{aligned}
&F(S_x, D_x, D_t)
\vartheta(\boldsymbol{\xi},\boldsymbol{\varepsilon'},
\boldsymbol{0}|\boldsymbol{\tau})\cdot
  \vartheta(\boldsymbol{\xi},\boldsymbol{\varepsilon}, \boldsymbol{0}|\boldsymbol{\tau})
  =\left[\sum_{\boldsymbol{\mu}}C(\boldsymbol{\varepsilon'},\boldsymbol{\varepsilon}, \boldsymbol{\mu})\right]
  \vartheta(2\boldsymbol{\xi},\boldsymbol{\varepsilon'}
  +\boldsymbol{\varepsilon}, \boldsymbol{\mu}/2|2\boldsymbol{\tau}),\end{aligned}\eqno(3.7)$$
in which, explicitly
$$\begin{aligned}
  &C(\boldsymbol{\varepsilon},\boldsymbol{\varepsilon'},\boldsymbol{\mu})=\sum_{\boldsymbol{n}\in \mathbb{Z}^N}
  F(\boldsymbol{\mathcal{M}})\exp\left[-2\pi\langle \boldsymbol{\tau}(\boldsymbol{n}-\boldsymbol{\mu}/2), \boldsymbol{n}
  -\boldsymbol{\mu}/2\rangle-2\pi i
  \langle \boldsymbol{n}-\boldsymbol{\mu}/2, \boldsymbol{\varepsilon'}-\boldsymbol{\varepsilon})\right].
\end{aligned}\eqno(3.8)$$
where we denote $ \boldsymbol{\mathcal{M}}=(4\pi i\langle
\boldsymbol{n}-\boldsymbol{\mu}/2, \boldsymbol{\alpha}\rangle,\ 4\pi
i\langle
  \boldsymbol{n}-\boldsymbol{\mu}/2, \boldsymbol{\omega}\rangle,\
  4\pi i\langle \boldsymbol{n} -\boldsymbol{\mu}/2, \boldsymbol{\sigma}+\theta\boldsymbol{\alpha}\rangle).$

{\it Proof.} For simplicity we prove the formula (3.6)  for
one-dimensional case. The proof for $N$-dimensional case can be
performed simply by replacing one-dimensional vectors by
$N$-dimensional ones.

Making use of  Proposition 1, we obtain the relation
\begin{eqnarray*}
&&\Delta\equiv S_x \vartheta(\xi,\varepsilon',0|{\tau})\cdot
  \vartheta(\xi,  \varepsilon, 0|{\tau})\\
  &&=\sum_{m',m\in\mathbb{Z}}\mathfrak{D}_x\exp\{2\pi i m'(\xi+\varepsilon')-\pi m'^2{\tau}\}\cdot
  \exp\{2\pi i m(\xi+\varepsilon)-\pi m^2{\tau}\},\\
  &&=\sum_{m',m\in\mathbb{Z}}2\pi i(\beta\sigma+\theta\alpha)(m'-m) \exp\left\{2\pi
  i(m'+m)\xi-2\pi i(m'\varepsilon'+m\varepsilon)-\pi{\tau}[m'^2+m^2]\right\}
  \end{eqnarray*}
  By shifting sum index as   $m=l'-m'$, then
  \begin{eqnarray*}
  &&\Delta=\sum_{l',m'\in\mathbb{Z}}2\pi i(\sigma+\theta\alpha)(2m'-l') \exp\left\{2\pi
  il'\xi-2\pi i[m'\varepsilon'+(l'-m')\varepsilon] -\pi{\tau}[m'^2+(l'-m')^2]\right\}\\
  &&\stackrel{l'=2l+\mu}{=}\sum_{\mu=0,1}\ \ \sum_{l,m'\in\mathbb{Z}}2\pi i(\beta\sigma+\theta\alpha)(2m'-2l-\mu)\exp\{4\pi
  i\xi(l+\mu/2)\\
  &&\ \ \ \ \   -2\pi i[m'\varepsilon'-(m'-2l-\mu)\varepsilon] -\pi[m'^2+(m'-2l-\mu)^2]{\tau}\}
  \end{eqnarray*}
Finally letting  $m'=n+l$,  we conclude  that
{\small\begin{eqnarray*}
    &&\Delta=\sum_{\mu=0,1}\left[\sum_{n\in\mathbb{Z}}4\pi i(\beta\sigma+\theta\alpha)[n-\mu/2]
    \exp\{-2\pi i(n-\mu/2)(\varepsilon'-\varepsilon)-2\pi{\tau}(n-\mu/2)^2\}\right]\\
&& \ \ \ \ \ \ \ \ \times \left[\sum_{l\in\mathbb{Z}}\exp\{2\pi
i(l+\mu/2)(2\xi+\varepsilon'+\varepsilon)-2\pi{\tau}(l+\mu/2)^2\right]\\
&&=\left[\sum_{\mu=0,1}\mathfrak{D}_x\vartheta(2\xi,\varepsilon'-\varepsilon,
-\mu/2|2{\tau})|_{\xi=0}\right]
  \vartheta(2\xi,\varepsilon'+\varepsilon, \mu/2|2\tau),
\end{eqnarray*}}
by using  the following relations
\begin{eqnarray*}
    &&n+l=(n-\mu/2)+(l+\mu/2),\ \ n-l-\mu=(n-\mu/2)-(l+\mu/2).
\end{eqnarray*}

In a similar way, we can prove the formula (3.5).  The formula (3.7)
follows from  (3.5) and (3.6). $\Box$

{\bf Remark 2.}  The formulae (3.7) and (3.8) show  that if the
following equations are   satisfied
  $$C(\boldsymbol{\varepsilon},\boldsymbol{\varepsilon'},\boldsymbol{\mu})=0,\eqno(3.9)$$
  for all possible combinations
  $\mu_1=0,1; \mu_2=0,1;  \cdots;  \mu_N=0,1$, in other word, all
  such  combinations  are solutions of equation (3.9),   then
  $\vartheta(\boldsymbol{\xi},\boldsymbol{\varepsilon'}, \boldsymbol{0}|
  \boldsymbol{\tau})$ and $
  \vartheta(\boldsymbol{\xi},\boldsymbol{\varepsilon}, \boldsymbol{0}|\boldsymbol{\tau})$
  are  $N$-periodic wave solutions of the bilinear equation
  $$F(S_t, D_x, D_t)\vartheta(\boldsymbol{\xi},\boldsymbol{\varepsilon'}, \boldsymbol{0}|
  \boldsymbol{\tau})\cdot\vartheta(\boldsymbol{\xi},\boldsymbol{\varepsilon}, \boldsymbol{0}|\boldsymbol{\tau})=0.$$
 We call the formula (3.9) constraint equations, whose number  is  $2^N$.
 This formula  actually provides us an unified approach to construct
  multi-periodic wave  solutions for supersymmetric equations. Once a
  supersymmetric equation is written bilinear forms, then its  multi-periodic wave  solutions
  can be directly obtained by solving system (3.9).

{\bf Theorem 2.} Let
$C(\boldsymbol{\varepsilon},\boldsymbol{\varepsilon'},\boldsymbol{\mu})$
and $F(S_x, D_x, D_t)$  be  given in Theorem 1,  and  make a choice
such that $\varepsilon_j'-\varepsilon_j=\pm 1/2, \ j=1, \cdots, N$.
Then

(i) \ If  $F(S_x, D_x, D_t)$ is an even function in the form
$$F(-S_x, -D_x, -D_t)=F(S_x, D_x, D_t),$$
 then $C(\boldsymbol{\varepsilon},\boldsymbol{\varepsilon'},\boldsymbol{\mu})$ vanishes automatically for
 the case when  $\sum_{j=1}^{N}\mu_j$ is
an odd number, namely
  $$C(\boldsymbol{\varepsilon},\boldsymbol{\varepsilon'},\boldsymbol{\mu})|_{\boldsymbol{\mu}}=0, \ \  {\rm for} \  \ \
 \sum_{j=1}^{N}\mu_j=1,\ {\rm mod }\ 2.\eqno(3.10)$$

(ii) \ If $F(S_x, D_x, D_t)$ is an odd function in the form
$$F(-S_x, -D_x,
-D_t)=-F(S_x, D_x, D_t),$$ then
$C(\boldsymbol{\varepsilon},\boldsymbol{\varepsilon'},\boldsymbol{\mu})$
vanishes automatically for
 the case when  $\sum_{j=1}^{N}\mu_j$ is
an even number, namely
$$C(\boldsymbol{\varepsilon},\boldsymbol{\varepsilon'},\boldsymbol{\mu})|_{\boldsymbol{\mu}}=0,
\ \ {\rm for} \ \sum_{j=1}^{N}\mu_j=0, \ {\rm mod }\ 2.\eqno(3.11)$$

{\it Proof.}  We are going to consider the case where  $F(S_x, D_x,
D_t)$ is an even function and prove the formula (3.9). The formula
(3.11) is analogous. Making transformation
$\boldsymbol{n}=-\boldsymbol{\bar{n}}+\boldsymbol{\mu} \ (
\boldsymbol{\bar{n}}=(\bar{n}_1, \cdots, \bar{n}_N),\
\bar{n}_j\in\mathbb{Z}, \ \ j=1, \cdots, N.$ ),  and noting $F(S_x,
D_x, D_t)$ is even, we then deduce that
$$\begin{aligned}
  &C(\boldsymbol{\varepsilon},\boldsymbol{\varepsilon'},\boldsymbol{\mu})
  =\sum_{\boldsymbol{\bar{n}}\in \mathbb{Z}^N}
  F(-\boldsymbol{\mathcal{M}})\exp\left[-2\pi\langle \boldsymbol{\tau}(\boldsymbol{\bar{n}}-\boldsymbol{\mu}/2), \boldsymbol{\bar{n}}
  -\boldsymbol{\mu}/2\rangle+2\pi i
  \langle \boldsymbol{\bar{n}}-\boldsymbol{\mu}/2, \boldsymbol{\varepsilon'}-\boldsymbol{\varepsilon})\right]\\
  &\ \ \ \ \ \ \ \ \ \ \ \ \   \ =C(\boldsymbol{\varepsilon},\boldsymbol{\varepsilon'},\boldsymbol{\mu})
  \exp\left(4\pi
  i\langle \boldsymbol{\bar{n}}-\boldsymbol{\mu}/2, \boldsymbol{\varepsilon'}-\boldsymbol{\varepsilon}\rangle\right)\\
  &\ \ \ \ \ \ \ \ \ \ \ \ \   \ =C(\boldsymbol{\varepsilon},\boldsymbol{\varepsilon'},\boldsymbol{\mu})
  \exp\left(\pm 2\pi i\sum_{j=1}^{N}\bar{n}_j\right)
  \exp\left(\pm\pi i\sum_{j=1}^{N}\mu_j\right)\\
  &\ \ \ \ \ \ \ \ \ \ \ \ \   \  =C(\boldsymbol{\varepsilon},\boldsymbol{\varepsilon'},\boldsymbol{\mu})\exp(\pm\pi i)
  =-C(\boldsymbol{\varepsilon},\boldsymbol{\varepsilon'},\boldsymbol{\mu}),
\end{aligned}$$
which proves  the formula (3.10). $\square$

{\bf Corollary 1. }  Let $\varepsilon_j'-\varepsilon_j=\pm 1/2, \
j=1, \cdots, N$.  Assume  $F(S_x, D_x, D_t)$ is a linear combination
of  even and  odd functions
$$F(S_x, D_x, D_t)=F_1(S_x, D_x, D_t)+F_2(S_x, D_x, D_t),$$
where $F_1(S_x, D_x, D_t)$ is  even  and $F_2(S_x, D_x, D_t)$ is
odd.  In addition,
$C(\boldsymbol{\varepsilon},\boldsymbol{\varepsilon'},\boldsymbol{\mu})$
corresponding (3.8)  is given by
$$C(\boldsymbol{\varepsilon},\boldsymbol{\varepsilon'},\boldsymbol{\mu})
=C_1(\boldsymbol{\varepsilon},\boldsymbol{\varepsilon'},\boldsymbol{\mu})
+C_2(\boldsymbol{\varepsilon},\boldsymbol{\varepsilon'},\boldsymbol{\mu}),$$
where
  $$C_1(\boldsymbol{\varepsilon},\boldsymbol{\varepsilon'},\boldsymbol{\mu})
=\sum_{\boldsymbol{n}\in \mathbb{Z}^N}
  F_1(\boldsymbol{\mathcal{M}})
  \exp\left[-2\pi\langle \boldsymbol{\tau}(\boldsymbol{n}-\boldsymbol{\mu}/2), \boldsymbol{n}
  -\boldsymbol{\mu}/2\rangle-2\pi i
  \langle \boldsymbol{n}-\boldsymbol{\mu}/2,
  \boldsymbol{\varepsilon'}-\boldsymbol{\varepsilon})\right],$$
  $$C_2(\boldsymbol{\varepsilon},\boldsymbol{\varepsilon'},\boldsymbol{\mu})
=\sum_{\boldsymbol{n}\in \mathbb{Z}^N}
  F_2(\boldsymbol{\mathcal{M}})
  \exp\left[-2\pi\langle \boldsymbol{\tau}(\boldsymbol{n}-\boldsymbol{\mu}/2), \boldsymbol{n}
  -\boldsymbol{\mu}/2\rangle-2\pi i
  \langle \boldsymbol{n}-\boldsymbol{\mu}/2,
  \boldsymbol{\varepsilon'}-\boldsymbol{\varepsilon})\right].$$
Then
$$\begin{aligned}
  &C(\boldsymbol{\varepsilon},\boldsymbol{\varepsilon'},\boldsymbol{\mu})
  =C_2(\boldsymbol{\varepsilon},\boldsymbol{\varepsilon'},\boldsymbol{\mu}) \ \  {\rm for} \  \ \
 \sum_{j=1}^{N}\mu_j=1,\ {\rm mod }\ 2,
\end{aligned}\eqno(3.12)$$
$$\begin{aligned}
  &C(\boldsymbol{\varepsilon},\boldsymbol{\varepsilon'},\boldsymbol{\mu})
  =C_1(\boldsymbol{\varepsilon},\boldsymbol{\varepsilon'},\boldsymbol{\mu}), \ \ {\rm for} \ \sum_{j=1}^{N}\mu_j=0,
   \ {\rm mod }\ 2.
\end{aligned}\eqno(3.13)$$

{\it Proof.}  In a similar to the proof of Theorem 2,  shifting sum
index as  $\boldsymbol{n}=-\boldsymbol{\bar{n}}+\boldsymbol{\mu}$,
and using $F_1(S_x, D_x, D_t)$ even and $F_2(S_x, D_x, D_t)$ odd, we
have
$$\begin{aligned}
  &C(\boldsymbol{\varepsilon},\boldsymbol{\varepsilon'},\boldsymbol{\mu})
  =C_1(\boldsymbol{\varepsilon},\boldsymbol{\varepsilon'},\boldsymbol{\mu})+
  C_2(\boldsymbol{\varepsilon},\boldsymbol{\varepsilon'},\boldsymbol{\mu})\\
    &\ \ \ \ \ \ \ \ \ \ \ \ \   \
    =\left[C_1(\boldsymbol{\varepsilon},\boldsymbol{\varepsilon'},\boldsymbol{\mu})-
  C_2(\boldsymbol{\varepsilon},\boldsymbol{\varepsilon'},\boldsymbol{\mu})\right]
  \exp\left(\pm\pi i\sum_{j=1}^{N}\mu_j\right).
\end{aligned}\eqno(3.14)$$

Then  for  $\sum_{j=1}^{N}\mu_j=1,\ {\rm mod }\ 2$,  the equation
(3.15) gives
$$C_1(\boldsymbol{\varepsilon},\boldsymbol{\varepsilon'},\boldsymbol{\mu})=0,$$
which implies the formula (3.12). The formula (3.13) is analogous.
$\square$

The theorem 2 and corollary 1 are very useful to deal with coupled
super-Hirota's bilinear equations, which will  be  seen  in the
following section 5.

By introducing differential operators
\begin{eqnarray*}
&&\nabla=(\partial_{\xi_1}, \partial_{\xi_2}, \cdots, \partial_{\xi_N}), \\
&&\partial_x=\alpha_1\partial_{\xi_1}+\alpha_2
\partial_{\xi_2}+\cdots+\alpha_N
\partial_{\xi_N}=\boldsymbol{\alpha}\cdot\nabla,\\
&&\partial_t=\omega_1\partial_{\xi_1}+\omega_2
\partial_{\xi_2}+\cdots+\omega_N
\partial_{\xi_N}=\boldsymbol{\omega}\cdot\nabla, \\
&&\mathfrak{D}=(\sigma_1+\theta\alpha_1)\partial_{\xi_1}+(\sigma_2+\theta\alpha_2)
\partial_{\xi_2}+\cdots+(\sigma_N+\theta\alpha_N)
\partial_{\xi_N}=(\boldsymbol{\sigma}+\theta\boldsymbol{\alpha})\cdot\nabla,
\end{eqnarray*}
then we have
 $$\begin{aligned}
 &\mathfrak{D}\partial_x^k\partial_t^j\vartheta(\boldsymbol{\xi},\boldsymbol{\tau})
 =[(\boldsymbol{\sigma}+\theta\boldsymbol{\alpha})\cdot\nabla](\boldsymbol{\alpha}\cdot\nabla)^k
 (\boldsymbol{\omega}\cdot\nabla)^j\vartheta(\boldsymbol{\xi},\boldsymbol{\tau})\\
 &=(\boldsymbol{\sigma}\cdot\nabla)(\boldsymbol{\alpha}\cdot\nabla)^k
 (\boldsymbol{\omega}\cdot\nabla)^j\vartheta(\boldsymbol{\xi},\boldsymbol{\tau})
 +\theta(\boldsymbol{\alpha}\cdot\nabla)^{k+1}
 (\boldsymbol{\omega}\cdot\nabla)^j\vartheta(\boldsymbol{\xi},\boldsymbol{\tau}),\\
 & j, k=0, 1, \cdots.
\end{aligned}\eqno(3.15)$$
\\[12pt]
%%%%%%%%%%%%%%%%%%%%%%%%%%%%%%%%%%%%%%%%%%%%%%%%%%%%%%%%%%%%%%%%%%%%%%%%%%%%%%%%%%%%%%%%%%%%%%%%%%%%%%%%%%%%%%%%%%%%%%%%%%%%%%%%%%%%%%
%%%%%%%%%%%%%%%%%%%%%%%%%%%%%%%%%%%%%%%%%%%%%%%%%%%%%%%%%%%%%%%%%%%%%%%%%%%%%%%%%%%%%%%%%%%%%%%%%%%%%%%%%%%%%%%%%%%%%%%%%%%%%%%%%%%%%%
{\bf\large 4.   $\mathcal{N}=1$ supersymmetric Sawada-Kotera-Ramani equation}\\

The  supersymmetric Sawada-Kotera-Ramani equation takes the form
$$\begin{aligned}
&\phi_t+\mathfrak{D}^2\left[10(\mathfrak{D}\phi)\mathfrak{D}^4\phi+5(\mathfrak{D}^5\phi)\phi
+15(\mathfrak{D}\phi)^2\phi\right]+\mathfrak{D}^{10}\phi=0,
\end{aligned}\eqno(4.1)$$
 where $\phi=\phi(x,t,\theta):\mathbb{R}_{\Lambda}^{2,1}\rightarrow \mathbb{R}_{\Lambda}^{0,1}$
 is fermionic superfield depending on usual independent variable $x$, $t$ and
 Grassmann variable $\theta$. This
 equation  was first proposed by Carstea \cite{ Carstea}.   The
 soliton solutions, Lax representation and infinite conserved quantities
 of the equation have been further obtained  recently \cite{Yu,
 Tian}. Here we are interested in quasi-periodic wave solutions to the
supersymmetric  equation (4.1).  We will show that the soliton
solutions can be obtained as limiting case of these quasi-periodic
solutions.

 To apply the Hirota bilinear method in superspace for constructing
multi-periodic wave solutions of the equation (4.1),  we hope more
fee  variables and  consider  a  general variable transformation
$$\phi=\phi_0+2\mathfrak{D}^3 \ln f(x, t, \theta),\eqno(4.2)$$
where  $f(x,t,\theta):\mathbb{R}_{\Lambda}^{2,1}\rightarrow
\mathbb{R}_{\Lambda}^{1,0}$ is an even superfield and
$\phi_0=\phi_0(\theta):\mathbb{R}_{\Lambda}^{2,1}\rightarrow
\mathbb{R}_{\Lambda}^{0,1}$ is an odd special solution of the
equation (4.1). Substituting (4.2) into (4.1) and integrating with
respect to $x$, we then get the following super Hirota's bilinear
form
$$\begin{aligned}
&F(S_x, D_x,D_t)f\cdot
f=(S_xD_t+S_xD_x^5+15\phi_0^2D_x^2+5\phi_0D_x^4+c) f\cdot f=0,
\end{aligned}\eqno(4.3)$$ where
 $c=c(\theta,t):\mathbb{R}_{\Lambda}^{2,1}\rightarrow \mathbb{R}_{\Lambda}^{0,1}$ is an
odd integration constant. In the special case when $\phi_0=c=0$,
starting from the bilinear equation (4.3), it is easy to find that
the equation (4.1) admits one-soliton solution (also called
one-supersoliton solution) in superspace $\mathbb{R}_\Lambda^{2,1}$
over two-dimensional Grassmann algebra $G_1(\sigma)$
$$\phi_1=2\mathfrak{D}^3\ln(1+e^{\eta}),\eqno(4.4)$$
with  phase variable $\eta= p x-p^5 t+q\theta \sigma+r$ with $p, q,
r\in\Lambda_0$. While  two-soliton solution (super two-soliton
solution)  reads
$$\phi_2=2\mathfrak{D}^3\ln(1+e^{\eta_1}+e^{\eta_2}+e^{\eta_1+\eta_2+A_{12}}),\eqno(4.5)$$
with $\eta_j=p_jx-p_j^3t+q_j\theta \sigma+r_j, \ \ j=1,2$ and
\begin{eqnarray*}
&&
e^{A_{12}}=\frac{(p_1-p_2)^2(p_1^2-p_1p_2+p_2^2)}{(p_1+p_2)^2(p_1^2+p_1p_2+p_2^2)}
\left(1+2\theta\sigma\frac{p_1q_2-p_2q_1}{p_1-p_2}\right),
\end{eqnarray*}
and here $p_j, q_j, r_j\in\Lambda_0, j=1, 2$ are free constants.

Next, we turn to see the periodicity of the solution (4.2), the
function  $f$ is chosen to be a Riemann theta function, namely,
 $$f(x,t,\theta)=\vartheta(\boldsymbol{\xi}, \boldsymbol{\tau}),$$
where  phase variable $\xi$  is taken as the form $\boldsymbol{\xi}
=(\xi_1,\cdots, \xi_N)^T$, $ \xi_j=\alpha_jx+\omega_j
t+\beta_j\theta\sigma +\delta_j, \ \ j=1,2, \cdots, N.$   With
Proposition 3, we refer to
$$\phi=\phi_0+2\sum_{k,l=1}^N\alpha_k(\beta_l\sigma+\theta\alpha_l)
\partial_{\xi_k\xi_l}^2\ln\vartheta(\boldsymbol{\xi},\boldsymbol{\tau}),$$
which shows that  the solution  $\phi$ is indeed  a quasi-periodic
function with $2N$ fundamental periods $\{\boldsymbol{e_j},\ \ j=1,
\cdots, N\}$ and $\{i\boldsymbol{\tau}_j, \ \ j=1, \cdots, N\}$. The
quasi-periodic means that $\phi$ is periodic in each of the $N$
phases $\{\xi_j, \ \ j=1,
\cdots, N\}$, if the other $N-1$ phases are fixed.\\[12pt]
%%%%%%%%%%%%%%%%%%%%%%%%%%%%%%%%%%%%%%%%%%%%%%%%%%%%%%%%%%%%%%%%%%%%%%%%%%%%%%%%%%%%%%%%%%%%%%%%%%%%%%%%%%%%%%%%%%%%%%%%%%%%%%%%%%%
{\bf 3.1. One-periodic waves and  asymptotic analysis}\\

 We first consider one-periodic wave solutions of the equation
 (4.1).  As a  simple case of  the theta function
 (3.1)  when $N=1, s=\varepsilon=0$,  we take $f$ as
$$f(x,t,\theta)=\vartheta(\xi,\tau)=\sum_{n\in\mathbb{Z}}\exp({2\pi in\xi-\pi n^2\tau}),\eqno(4.6)$$
where the phase variable $\xi=\alpha x+\omega
t+\beta\theta\sigma+\delta$, and the parameter $\tau>0$.

Next, we  let the Riemann theta function (4.6) be a solution of the
bilinear equation (4.3).   By using Theorem 1 and the formula (3.9),
the following two equations ( corresponding to $\mu=0$ and $1$
respectively) should be satisfied
 $$\begin{aligned}
&\sum_{n\in\mathbb{Z}}\{-(4\pi(
n-\mu/2))^2(\beta\sigma+\theta\alpha)\omega-(4\pi(
n-\mu/2))^6(\beta\sigma+\theta\alpha)\alpha^5-15(4\pi(n-\mu/2))^2\alpha^2\phi_0^2\\
&+5(4\pi( n-\mu/2))^4\alpha^4\phi_0+c\}\exp(-2\pi \tau
(n-\mu/2)^2)=0, \ \ \mu=0, 1.
\end{aligned}\eqno(4.7)$$

We introduce the notations by
$$\begin{aligned}
&\lambda=e^{-\pi\tau/2 },\quad
\vartheta_1(\xi,\lambda)=\vartheta(2\mathbf{\xi},0,0|
2\tau)=\sum_{n\in\mathbb{Z}}
\lambda^{4n^2}\exp(4i\pi n\xi),\\
&\vartheta_2(\xi,\lambda)=\vartheta(2\xi,0,
-1/2|2\tau)=\sum_{n\in\mathbb{Z}} \lambda^{(2n-1)^2}\exp[2i\pi(2n-1)
\xi],\end{aligned})$$
 the equation (4.7) can be written as
a linear system about $\omega, c$
 $$\begin{aligned}
&(\beta\sigma+\theta\alpha)\vartheta_j''\omega+(\beta\sigma+\theta\alpha)\alpha^5\vartheta_j^{(6)}
+3\alpha^2\vartheta_j''\phi_0^2+5\alpha^4\vartheta_j^{(4)}\phi_0+\vartheta_jc=0,\
\ j=1,\ 2,
\end{aligned}\eqno(4.8)$$
where $\omega\in\Lambda_0$ is even and $c, \phi_0\in\Lambda_1$ are
odd, and we have denoted the derivative value of
$\vartheta_j(\xi,\lambda)$ at $\xi=0$ by simple notations
$$\vartheta_j'=\vartheta_j'(0,\lambda)=\frac{d\vartheta_j(\xi,\lambda)}{d\xi}|_{\xi=0}, \ \ j=1,2.$$
Moreover, we see that the functions $\vartheta_j$ and their
derivatives are independent of Grassmann variable $\theta$ and
anticommuting number $\sigma$.

We take $\phi_0=0$ for the simplicity. It is obvious that the
coefficient determinant of the system (4.8) is nonzero and
$(\alpha^5\vartheta_1^{(6)}, \alpha^5\vartheta_2^{(6)})^T\not=0$,
therefore the system (4.8) admits a solution
 $$\begin{aligned}
&\omega=\frac{\alpha^5(\vartheta_2^{(6)}\vartheta_1-\vartheta_1^{(6)}\vartheta_2)}
{\vartheta_1''\vartheta_2-\vartheta_2''\vartheta_1},\ \ \
b_1=\frac{\alpha^5(\beta\sigma+\theta\alpha)(\vartheta_2^{(6)}\vartheta_1''-\vartheta_1^{(6)}\vartheta_2'')}
{\vartheta_1''\vartheta_2-\vartheta_2''\vartheta_1},
\end{aligned}\eqno(4.9)$$
where $\omega$ is independent of Grassmann variable $\theta$ and
auticommuting number $\sigma$, and  parameter $\alpha$ is free.

 In this way,  a one-periodic wave solution of the equation (4.1) is
explicitly obtained by
$$\phi=2\mathfrak{D}^3\ln \vartheta(\xi,\tau),\eqno(4.10)$$
with the theta function  $\vartheta(\xi,\tau) $ given by (4.6) and
parameters $\omega$, $c$ by (4.9), while  other parameters $\alpha,
\beta, \tau, \delta\in\Lambda_0$ are free. Among them, the three
parameters $\alpha$ and $\tau$ completely dominate a one-periodic
wave.

In summary,  one-periodic wave (4.10) possesses the following
features:

 (i) It is one-dimensional, i.e. there is a single phase
variable $\xi$. Moreover, it  has two fundamental periods $1$ and
$i\tau$ in phase variable $\xi$, but it need not to be periodic in
$x$, $t$ and $\theta$ directions.

(ii) It can be viewed as a parallel superposition of overlapping
one-soliton waves, placed one period apart ( see $(a)$ and $(b)$ in
Figure 1 ).

(iii) Different form the purely bosonic case,  it is observed shows
that there is an influencing band among the one-periodic waves under
the presence of the Grassmann variable (in contour plot, the bright
hexagons are crests and the dark hexagons are troughs). The
one-periodic waves are symmetric about the band but collapse along
with the band. Furthermore, the amplitudes of the quasi-periodic
waves increase as the waves move away from the band ( see $(a)$ and
$(b)$ in Figure 1 ).

(iv) The quasi-periodic wave will degenerate to pure bosonic
quasi-periodic wave  when $\theta$ becomes small ( see Figure 2 ).

\input epsf
\begin{figure}
\centering {\footnotesize $(b) \ \ \ \ \ \ \ \ \ \   \ \ \ \ \ \ \ \
\ \ \ \ \  \ \ \ \ \ \  $}\  \ \  \ \ \ \ \ \ \ \ \ \ \ \
 \ \ {\footnotesize $(b)$}$ \ \ \ \ \ \ \  \ \ \ \
 \ \ \ \ $\\
\includegraphics[width=2.1 in,angle=0]{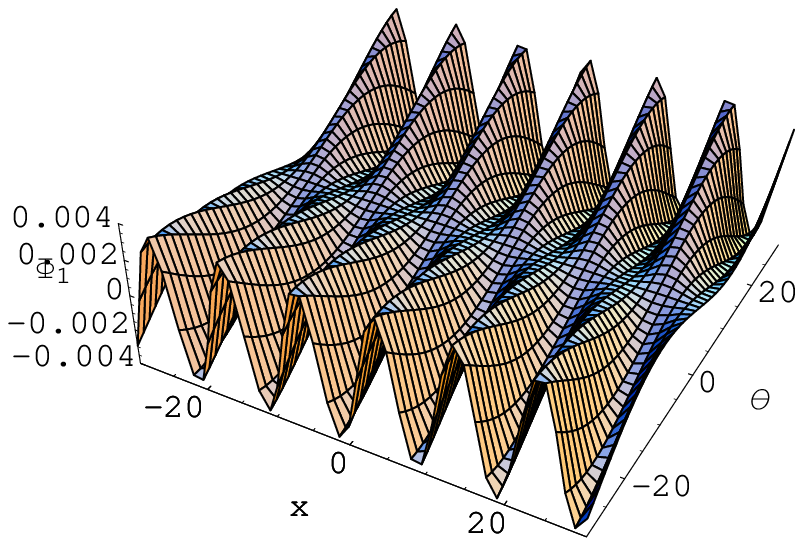}\ \ \ \ \ \  \ \ \ \
\ \
\includegraphics[width=1.4 in,angle=0]{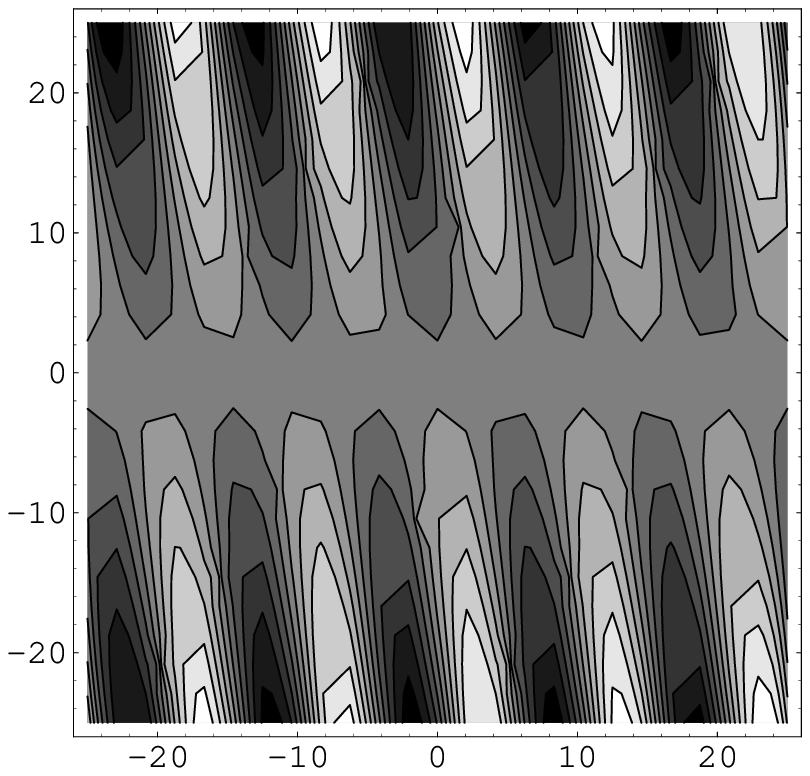}
\\[6pt]
\caption{\footnotesize\baselineskip=15pt A  one-periodic wave for
$\mathcal{N}=1$ supersymmetric Sawada-Kotera-Ramani equation with
parameters: $\alpha=0.1,$ $\tau=32, \sigma=0.013$.  (a) Perspective
view of wave. (b) Overhead view of wave, with contour plot shown.}
\end{figure}

\input epsf
\begin{figure}
\centering {\footnotesize $(b) \ \ \ \ \ \ \ \ \ \   \ \ \ \ \ \ \ \
\ \ \ \ \  \ \ \ \ \ \  $}\  \ \  \ \ \ \ \ \ \ \ \ \ \ \
 \ \ {\footnotesize $(b)$}$ \ \ \ \ \ \ \  \ \ \ \
 \ \ \ \ $\\
\includegraphics[width=2.1 in,angle=0]{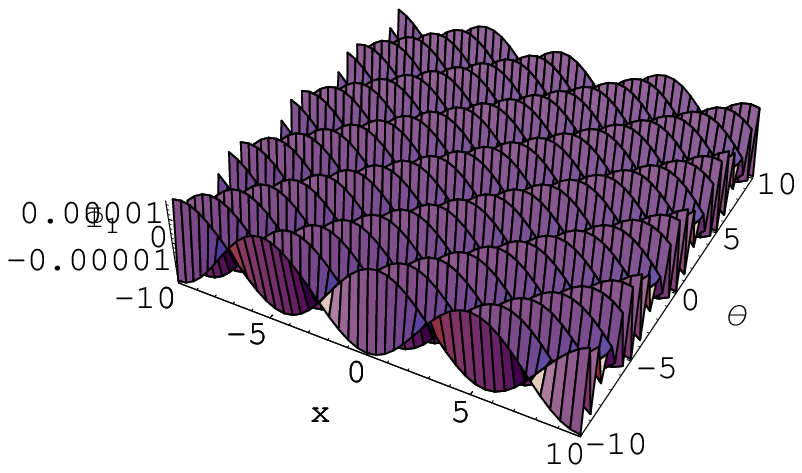}\ \ \ \ \ \  \ \ \ \
\ \
\includegraphics[width=1.4 in,angle=0]{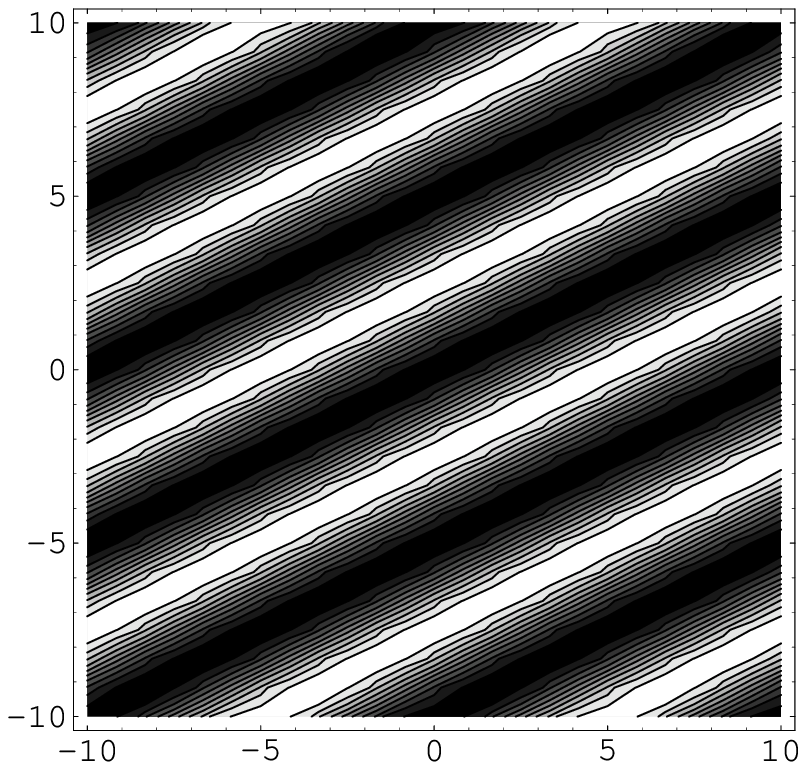}
\\[6pt]
\caption{\footnotesize\baselineskip=15pt  A purely bosonic case of
one-periodic wave to  $\mathcal{N}=1$ supersymmetric
Sawada-Kotera-Ramani equation with parameters: $\alpha=0.1,$ \
$\tau=32, \ \sigma=0.013, \ \theta=0.00000001 $.
 (a) Perspective view of wave. (b) Overhead view of wave, with contour plot shown.  }
\end{figure}

In the following, we further consider asymptotic properties of the
one-periodic wave solution. Interestingly,  the relation between the
one-periodic wave solution (4.10) and the one-soliton solution (4.4)
can be   established  as follows.

{\bf Theorem 3.}  Suppose that  $\omega$ and $c$ are given by (4.9),
and for the one-periodic wave solution (4.10), we let
$$\alpha=\frac{p}{2\pi i}, \ \ \beta=\frac{q}{2\pi i}, \ \
\delta=\frac{r+\pi \tau}{2\pi i},\eqno(4.11)$$ where the $p, q$ and
$r$ are given  in (4.4). Then we have the following asymptotic
properties
$$
\begin{aligned}
&c\longrightarrow 0, \ \  2\pi
i\xi-\pi\tau\longrightarrow\eta=px-p^5t
+q\theta\sigma+r,\\
&\vartheta(\xi,\tau)\longrightarrow 1+e^{\eta}, \ \ {\rm as } \ \
\lambda\rightarrow 0.
 \end{aligned}
 $$
 In
other words, the one-periodic solution (4.10)  tends to the soliton
solution (4.4) under a small amplitude limit, that is,
$$\phi\longrightarrow \phi_1, \ \ {\rm as } \ \
\lambda\rightarrow 0.\eqno(4.12)$$

{\it Proof.}  Here we will directly use the system (4.8) to analyze
asymptotic properties of one-periodic solution (4.10), which is more
simple and effective than  our original  method by solving the
system \cite{Dai}-\cite{Fan2}.  Since the coefficients of system
(4.8) are power series about $\lambda$, its solution $(\omega, c)$
also should be a series about $\lambda$.  We explicitly expand the
coefficients of system (4.8) as follows
$$\begin{aligned}
&\vartheta_1=1+2\lambda^4+\cdots,\quad \
\vartheta_1''=-32\pi^2\lambda^{4}+\cdots,
\\
&\vartheta_1^{(6)}=-8192\pi^6\lambda^4+\cdots, \ \ \vartheta_2=2\lambda+2\lambda^9+\cdots\\
&\vartheta_2''=-8\pi^2\lambda+\cdots,\ \ \
\vartheta_1^{(6)}=-128\pi^6\lambda+\cdots.\end{aligned}\eqno(4.13)$$
Let  the solution of the system (4.8) be of the form
$$\begin{aligned}
&\omega=\omega_0+\omega_1\lambda+\omega_2\lambda^2+\cdots=\omega_0+o(\lambda),\\
&c=c_0+c_1\lambda+c_2\lambda^2+\cdots=c_0+o(\lambda).
\end{aligned}\eqno(4.14)$$

Substituting the expansions (4.13) and (4.14) into the system (4.8)
(the second equation is divided by $\lambda$ ) and letting
$\lambda\longrightarrow 0$, we immediately obtain
 the following relations
$$
 \begin{aligned}
 &c_0=0, \ \ -8\pi^2(\beta\sigma+\theta\alpha)\omega_0+2c_0-128\pi^6(\beta\sigma+\theta\alpha)\alpha^5=0,
 \end{aligned}$$
which has a solution
 $$c_0=0, \ \ w_0=-16\pi^4\alpha^5.\eqno(4.15)$$
Combining (4.14) and (4.15) then  yields
$$c\longrightarrow 0, \ \  2\pi i\omega\longrightarrow -32i\pi^5\alpha^5=
-p^5, \ \ {\rm as } \ \ \lambda\rightarrow 0.$$ Hence we conclude
$$\begin{aligned}
&\hat{\xi}=2\pi i\xi-\pi \tau=p x+2\pi i\omega t+q\theta\sigma+r\\
&\quad \longrightarrow px-p^5t+q\theta\sigma+r=\eta,\ \ {\rm as}\ \
\lambda\rightarrow 0.
\end{aligned}\eqno(4.16)$$

 It remains to consider  asymptotic properties of  the one-periodic wave solution (4.10) under the limit
$\lambda\rightarrow 0$. By expanding the Riemann theta function
$\vartheta(\xi, \tau)$ and by using (4.16), it follows that
$$\begin{aligned}
&\vartheta(\xi,\tau)=1+\lambda^2(e^{2\pi i\xi}+e^{-2\pi
i\xi})+\lambda^8(e^{4\pi i\xi}+e^{-4\pi i\xi}) +\cdots
\\
&\ \ \ \
 \ =1+e^{\hat{\xi}}+\lambda^4(e^{-\hat{\xi}}+e^{2\hat{\xi}})+\lambda^{12}(e^{-2\hat{\xi}}+e^{3\hat{\xi}})
+\cdots\\
&\ \ \ \
 \ \quad \longrightarrow 1+e^{\hat{\xi}}\longrightarrow 1+e^{\eta},\ \
{\rm as}\ \ \lambda\rightarrow 0,
\end{aligned}$$
 which together with (4.10) lead to (4.12). Therefore we conclude that the one-periodic solution
(4.10) just goes to the one-soliton solution (4.4)  as the amplitude
$\lambda\rightarrow 0$. $\square$\\[12pt]
%%%%%%%%%%%%%%%%%%%%%%%%%%%%%%%%%%%%%%%%%%%%%%%%%%%%%%%%%%%%%%%%%%%%%%%%%%%%%%%%%%%%%%%%%%%%%%%%%%%%%%%%%%%%%%%%%%%%%%%%%%%%%%%%%%%%%%
%%%%%%%%%%%%%%%%%%%%%%%%%%%%%%%%%%%%%%%%%%%%%%%%%%%%%%%%%%%%%%%%%%%%%%%%%%%%%%%%%%%%%%%%%%%%%%%%%%%%%%%%%%%%%%%%%%%%%%%%%%%%%%%%%%%%%%
{\bf 3.2. Two-periodic wave solutions and asymptotic analysis }\\

We proceed to the construction of the  two-periodic wave solutions
to  the supersymmetric Sawada-Kotera-Ramani equation (4.1),  which
are a two-dimensional generalization of one-periodic wave solutions.
The two-periodic waves of interest here have three-dimensional
velocity fields and two-dimensional surface patterns.

For the case when $N=2,
\boldsymbol{s}=\boldsymbol{\varepsilon}=\boldsymbol{0}$ in the
Riemann theta function (3.1), we takes $f$ as
 $$f=\vartheta(\boldsymbol{\xi}, \boldsymbol{\tau})=\sum_{\boldsymbol{n}\in \mathbb{Z}^2}
 \exp\{2\pi i\langle\boldsymbol{\xi},\boldsymbol{n}\rangle-\pi
\langle\boldsymbol{\tau}
\boldsymbol{n},\boldsymbol{n}\rangle\},\eqno(4.17)$$ where
$\boldsymbol{n}=(n_1, n_2)^T\in \mathbb{Z}^2,\ \
\boldsymbol{\xi}=(\xi_1, \xi_2)^T\in \mathbb{C}^2,\ \
\xi_i=\alpha_jx+\omega_jt+\beta_j\theta\sigma+\delta_j, \ \ j=1, 2$;
The
 matrix $\boldsymbol{\tau}$ is a  positive definite  and  real-valued
symmetric ${2\times 2}$ matrix, which can  takes of the form
$$\boldsymbol{\tau}=(\tau_{ij})_{2\times 2}, \ \ \tau_{12}=\tau_{21}, \ \  \tau_{11}>0,\ \
 \tau_{22}>0,\ \ \tau_{11}\tau_{22}-\tau_{12}^2>0.$$

 Next, we explore  the conditions  to  make  the Riemann theta function (4.17) satisfy the bilinear equation
(4.3).   Theorem 1 and the formula (3.9) give rise to  the following
four  constraint  equations
  $$\begin{aligned}
&\sum_{n_1,n_2\in\mathbb{Z}}\left[-4\pi^2\langle
2\boldsymbol{n}-\boldsymbol{\mu},\boldsymbol{\sigma}+\theta\boldsymbol{\alpha}\rangle\langle
2\boldsymbol{n}-\boldsymbol{\mu},
\boldsymbol{\omega}\rangle-64\pi^6\langle
2\boldsymbol{n}-\boldsymbol{\mu},\boldsymbol{\alpha}\rangle^5\langle
2\boldsymbol{n}-\boldsymbol{\mu},\boldsymbol{\sigma}+\theta\boldsymbol{\alpha}\rangle\right.\\
&-60\pi^2\langle
2\boldsymbol{n}-\boldsymbol{\mu},\boldsymbol{\alpha}\rangle^2\phi_0^2+80\pi^4\langle
2\boldsymbol{n}-\boldsymbol{\mu},\boldsymbol{\alpha}\rangle^4\phi_0+c]\exp\{-2\pi
\langle \boldsymbol{\tau} (\boldsymbol{n}-\boldsymbol{\mu}/2),
\boldsymbol{n}-\boldsymbol{\mu}/2\rangle\}=0,
\end{aligned}\eqno(4.18)$$
where $\boldsymbol{\mu}=(\mu_1, \mu_2)$ takes all possible
combinations of $\mu_1, \mu_2=0,1$.

By introducing the notations
 \begin{eqnarray*}
 &&\lambda_{kl}=e^{-\pi\tau_{kl}/2}, k, l=1,2, \boldsymbol{\lambda}=(\lambda_{11},\lambda_{12},
\lambda_{22})\\
&&\vartheta_j(\boldsymbol{\xi},\boldsymbol{\lambda})=\vartheta(2\xi,\boldsymbol{0},-\boldsymbol{s}_j/2|2\tau)
 =\sum_{n_1,n_2\in Z}\exp\{4\pi i\langle\boldsymbol{\xi} ,\boldsymbol{n}-\boldsymbol{s_j}/2\rangle\}
 \prod_{k,l=1}^{2}\lambda_{kl}^{(2n_k-s_{j,k})(2n_j-s_{j,l})} ,\\
&& \boldsymbol{s_j}=(s_{j,1}, s_{j,2}),\quad j=1, 2, 3, 4,\ \
\boldsymbol{s_1}=(0,0),\ \ \boldsymbol{s_2}=(1,0),\ \
\boldsymbol{s_3}=(0,1),\ \ \boldsymbol{s_4}=(1,1),
\end{eqnarray*}
then by using (3.15), the  system (4.18) can be written as a linear
system
 $$\begin{aligned}
 &[(\boldsymbol{\beta}\sigma+\theta\boldsymbol{\alpha})\cdot\nabla]
 (\boldsymbol{\omega}\cdot\nabla)\vartheta_j(0,\boldsymbol{\lambda})+
 [(\boldsymbol{\beta}\sigma+\theta\boldsymbol{\alpha})\cdot\nabla]
 (\boldsymbol{\alpha}\cdot\nabla)^5\vartheta_j(0,\boldsymbol{\lambda})\\
 &+15(\boldsymbol{\alpha}\cdot\nabla)^2\vartheta_j(0,\boldsymbol{\lambda})\phi_0^2
+5(\boldsymbol{\alpha}\cdot\nabla)^4\vartheta_j(0,\boldsymbol{\lambda})\phi_0
 +\vartheta_j(0,\boldsymbol{\lambda})c=0.
 \end{aligned}\eqno(4.19)$$
  This system is easy to be solved in
such a way: $\phi_0$ by solving a quadratic equation with one
unknown; $\omega_1, \omega_2$ and $c$ by solving a linear system.
With such a solution $(\omega_1, \omega_2, \phi_0, c)$,  we then get
an exact two-periodic wave solution
$$\phi=\phi_0+2\mathfrak{D}^3\ln \vartheta(\boldsymbol{\xi}, \boldsymbol{\tau}),\eqno(4.20)$$
with $\vartheta(\boldsymbol{\xi},\boldsymbol{\tau})$ and $\omega_1,
\omega_2, \phi_0, c$ given by (4.17) and (4.19), respectively, while
other parameters $\beta_{1}, \beta_{2}, \alpha_1,$  $ \alpha_2,
\tau_{11}, \tau_{22}, \tau_{12}, \delta_1, \delta_2\in\Lambda_0$ are
free.

 In summary, two-periodic wave (4.20), which is a direct generalization
of one-periodic wave, has the following features:

(i) The two-periodic wave solution is genuinely
 two-dimensional. Its surface pattern is two-dimensional, namely, there are two
phase variables $\xi_1$ and $\xi_2$.

(ii) It has two independent spatial periods in two independent
horizontal directions.  It has $4$ fundamental periods $\{e_1,
e_2\}$ and $\{i\tau_1, i\tau_2\}$ in $(\xi_1, \xi_2)$. It is
spatially periodic in two directions $\xi_1, \xi_2$,
 but it does not need   periodic in the all  $x$-,  $t$-  and  $\theta$-directions.

(iii)  As in the case of on-periodic waves, there is an influencing
band among the two-periodic waves under the presence of the
Grassmann variable.  ( see  Figure 3 ).

\input epsf
\begin{figure}
\centering {\footnotesize $(a) \ \ \ \ \ \ \ \ \ \   \ \ \ \ \ \ \ \
\ \ \ \ \  \ \ \ \ \ \  $}\  \ \  \ \ \ \ \ \ \ \ \ \ \ \
 \ \ {\footnotesize $(b)$}$ \ \ \ \ \ \ \  \ \ \ \
 \ \ \ \ $\\
\includegraphics[width=2.1 in,angle=0]{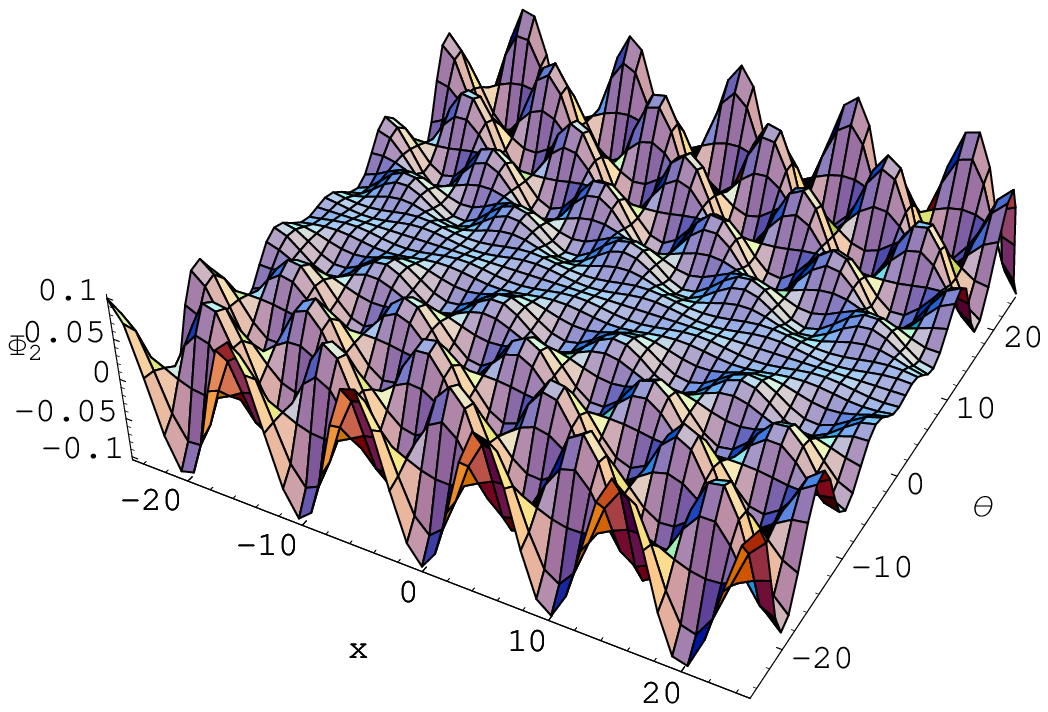}\ \ \ \ \ \  \ \ \ \
\ \
\includegraphics[width=1.4 in,angle=0]{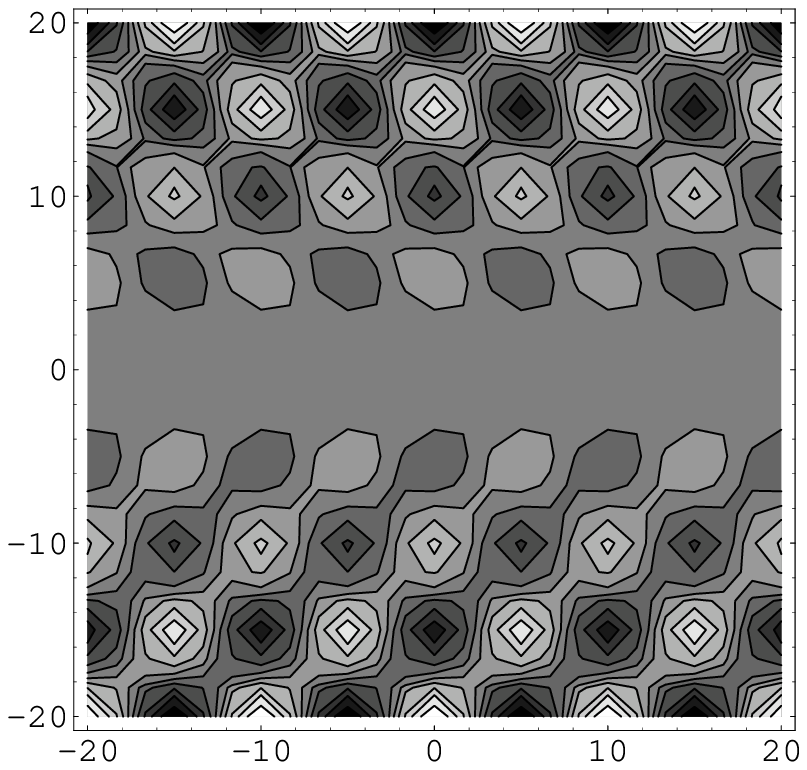}

\caption{\footnotesize\baselineskip=15pt A  two-super periodic wave
for $\mathcal{N}=1$ supersymmetric Sawada-Kotera-Ramani equation
with parameters: $\alpha_1=0.1, \alpha_2=-0.1, $ $\tau_{11}=2,
\tau_{12}=0.2, \tau_{22}=2, \sigma_1=0.1, \sigma_2=0.1$.   (a)
Perspective view of wave. (b) Overhead view of wave, with contour
plot shown. }
\end{figure}

 At last, we consider the asymptotic properties of the two-periodic solution (4.20).
 In a similar way to Theorem 3, we can
establish the relation between the two-periodic solution (4.20) and
the two-soliton solution as follows.

{\bf Theorem 4.}   Assume that $(\omega_1, \omega_2, \phi_0, c)^T$
is a solution of the system (4.19). We choose the parameters in  the
two-periodic wave solution (4.20)  as follows
$$
\begin{aligned}
&\alpha_j=\frac{p_j}{2\pi i},\ \ \beta_j=\frac{q_j}{2\pi i},\ \
\delta_j=\frac{r_j+\pi\tau_{jj} }{2\pi i},\ \
\tau_{12}=-\frac{A_{12}}{2\pi},\ \ j=1,2,
\end{aligned}\eqno(4.21)$$
 with the $p_j, q_j, r_j,
j=1, 2$ as those given in (4.5). Then under constraint
$\alpha_1\beta_2=\alpha_2\beta_1$, we have the following asymptotic
relations
$$
\begin{aligned}
&\phi_0\longrightarrow 0, \ \ \ c\longrightarrow 0, \ \ 2\pi
i\xi_j-\pi \tau_{jj}\longrightarrow
p_jx-p_j^5t+q_j\theta\sigma+r_j=\eta_j, \ \
j=1, 2,\\
& \vartheta(\boldsymbol{\xi}, \boldsymbol{\tau})\longrightarrow
1+e^{\eta_1}+e^{\eta_2}+e^{\eta_1+\eta_2+A_{12}}, \  \ \ {\rm as} \
\ \lambda_{11}, \lambda_{22}\rightarrow 0.
\end{aligned}\eqno(4.22)$$
So the two-periodic wave  solution (5.20) tends to the two-soliton
solution (4.5), namely,
$$\phi\longrightarrow \phi_2,
\ \ {\rm as }\ \ \lambda_{11}, \lambda_{22}\rightarrow 0.$$

{\it Proof.}  From (4.21), the constraint
$\alpha_1\beta_2=\alpha_2\beta_1$ leads to  $p_1q_2-p_2q_1=0$, which
implies that $\tau_{12}=-A_{12}/2\pi$ is independent of Grassmann
variable $\theta$ according to (4.5).

  In the same way as the proof of Theorem 3, we expand  the Riemann
  function $\vartheta(\xi_1, \xi_2, \boldsymbol{\tau})$ in the following form
\begin{eqnarray*}
&& \vartheta(\boldsymbol{\xi},\boldsymbol{\tau})=1+(e^{2\pi
i\xi_1}+e^{-2\pi i\xi_1})e^{-\pi {\tau}_{11}} +(e^{2\pi
i\xi_2}+e^{-2\pi i\xi_2})e^{-\pi
\tau_{22}}\\
&&\ \ \ \ \ \ \ \ \ \ \ \ +(e^{2\pi i(\xi_1+\xi_2)}+e^{-2\pi
i(\xi_1+\xi_2)})e^{-\pi (\tau_{11}+2\tau_{12}+\tau_{22})}+\cdots
\end{eqnarray*}
Further by using (4.21) and making a transformation
$\hat{\omega}_j=2\pi i \omega_j, j=1, 2$, we get
\begin{eqnarray*}
&&\vartheta(\boldsymbol{\xi},\boldsymbol{\tau})=1+e^{\hat{\xi}_1}+e^{\hat{\xi}_2}+e^{\hat{\xi}_1+\hat{\xi}_2-2\pi
\tau_{12} }+\lambda_{11}^4e^{-\hat{\xi}_1}
 +\lambda_{22}^4e^{-\hat{\xi}_2}+\lambda_{11}^4\lambda_{22}^4e^{-\hat{\xi}_1-\hat{\xi}_2-2\pi \tau_{12}}+\cdots\\
&& \ \ \ \ \ \longrightarrow
1+e^{\hat{\xi}_1}+e^{\hat{\xi}_2}+e^{\hat{\xi}_1+\hat{\xi}_2+A_{12}},\
\ {\rm as}\ \ \lambda_{11}, \lambda_{22} \rightarrow 0,
 \end{eqnarray*}
where $\hat{\xi}_j=p_jx+\hat{\omega}_jt+p_j\theta\sigma+r_j, \ \
j=1,2.$

 It remains to prove that
$$
\begin{aligned}
&c\longrightarrow 0, \ \ \ \hat{\omega}_j\longrightarrow-p_j^5, \ \
\hat{\xi}_j\longrightarrow \eta_j,\ \ j=1, 2, \ \ \ {\rm as} \ \
\lambda_{11}, \lambda_{22}\rightarrow 0.\end{aligned}\eqno(4.23)$$

As  in the case when $N=1$,  we let  the solution of the system
(4.19) be the form
$$\begin{aligned}
&\omega_1=\omega_{1,0}+\omega_{1,1}\lambda_{11}+\omega_{1,2}\lambda_{22}+\cdots=\omega_{1,0}+o(\lambda_{11},\lambda_{22}),\\
&\omega_2=\omega_{2,0}+\omega_{2,1}\lambda_{11}+\omega_{2,2}\lambda_{22}+\cdots=\omega_{2,0}+o(\lambda_{11},\lambda_{22}),\\
&\phi_0=\phi_{0,0}+\phi_{0,1}\lambda_{11}+\phi_{0,2}\lambda_{22}+\cdots=\phi_{0,0}+o(\lambda_{11},\lambda_{22}),\\
&c=c_0+c_1\lambda_{11}+c_2\lambda_{22}+\cdots=c_0+o(\lambda_{11},\lambda_{22}).
\end{aligned}\eqno(4.24)$$
Expanding  functions  $\vartheta_j, j=1,2,3,4$ in the system (4.19),
together with substitution of assumption (4.24), the second and
third equation is divided by $\lambda_{11}$ and  $\lambda_{22}$,
respectively; the fourth equation is divided by
$\lambda_{11}\lambda_{22}$, and letting
$\lambda_{11},\lambda_{22}\longrightarrow 0$ , we then obtain
$$\begin{aligned}
&c_0=0, \ -8\pi(\beta_1\sigma+\theta\alpha_1)\omega_1-128
\pi^6\alpha_1^5(\beta_1\sigma+\theta\alpha_1)-120\pi\alpha_1^2\phi_{0,0}^2+2\phi_{0,0}=0,\\
& -8\pi(\beta_2\sigma+\theta\alpha_2)\omega_2-128
\pi^6\alpha_2^5(\beta_2\sigma+\theta\alpha_2)-120\pi\alpha_2^2\phi_{0,0}^2+2\phi_{0,0}=0,
\end{aligned}$$
which has solution
$$
 \begin{aligned}
 &c_0=0, \ \ \phi_{0,0}=0, \ \ \ \omega_{1,0}=-16\pi^4\alpha_1^5, \ \
 \omega_{2,0}=-16\pi^4\alpha_2^5.
 \end{aligned}\eqno(4.25)$$
The expressions (4.24) and (4.25) implies  that
$$\begin{aligned}
&\phi_0=o(\lambda_{11}, \lambda_{22})\longrightarrow 0, \ \
c=o(\lambda_{11}, \lambda_{22})\longrightarrow 0,\ \
\omega_1=-16\pi^4\alpha_1^5+o(\lambda_{11},
\lambda_{22})\longrightarrow
-16\pi^4\alpha_1^5,\\
 &\omega_2=-16\pi^4\alpha_2^5+o(\lambda_{11}, \lambda_{22})\longrightarrow
-16\pi^4\alpha_2^5,  \ \ \ {\rm as } \ \ \lambda_{11},
\lambda_{22}\rightarrow 0,
\end{aligned}$$
thus  proving  (4.23).  We conclude that the two-periodic wave
solution (4.20) tends to the two-soliton solution (4.5)  as
$\lambda_{11}, \lambda_{22}\rightarrow 0$. $\square$\\ [12pt]
%%%%%%%%%%%%%%%%%%%%%%%%%%%%%%%%%%%%%%%%%%%%%%%%%%%%%%%%%%%%%%%%%%%%%%%%%%%%%%%%%%%%%%%%%%%%%%%%%%%%%%%%%
%%%%%%%%%%%%%%%%%%%%%%%%%%%%%%%%%%%%%%%%%%%%%%%%%%%%%%%%%%%%%%%%%%%%%%%%%%%%%%%%%%%%%%%%%%%%%%%%%%%%%%%%%
{\bf\large 4. $\mathcal{N}=2$ supersymmetric KdV equation}\\

We consider $\mathcal{N}=2$  supersymmetric KdV equation
$$\begin{aligned}
&\phi_t=-\phi_{xxx}+3(\phi\mathfrak{D}_1\mathfrak{D}_2\phi)_x+\frac{1}{2}(a-1)(\mathfrak{D}_1\mathfrak{D}_2\phi^2)_x
+3a\phi^2\phi_x^2,
\end{aligned}\eqno(5.1)$$
which was originally introduced by Laberge and Mathieu \cite{Lab1,
Lab2}. In the equation (5.1),  $\phi=\phi(x,t,\theta_1,
\theta_2):\mathbb{R}_{\Lambda}^{2,2}\rightarrow
\mathbb{R}_{\Lambda}^{0,1}$ is a superboson function depending on
temporal variable $t$, spatial variable $x$ and its fermionic
counterparts $\theta_1, \theta_2$. The operators $\mathfrak{D}_1$
and $\mathfrak{D}_2$ are the super derivatives defined by $
\mathfrak{D}_1=\partial_{\theta_1}+\theta_1\partial_x, \ \
\mathfrak{D}_2=\partial_{\theta_2}+\theta_2\partial_x $ and $a$ is a
parameter. The equation (5.1) is called supersymmetric KdV$_a$
equation \cite{Zhang}. For the cases when $a=1$ and $a=4$, the Lax
representation, Hamiltonian structure, Painleve analysis and soliton
solutions of the equation (5.1) can refer to, for instance,
  papers \cite{Lab1}--\cite{Zhang}.

Here we are interested in quasi-periodic wave solutions to the
supersymmetric  equation (5.1) by using Theorem 1 and 5. We only
consider the case when $a=1$, so the equation (5.1) reduces to
$$\begin{aligned}
&\phi_t=-\phi_{xxx}+3(\phi\mathfrak{D}_1\mathfrak{D}_2\phi)_x+3\phi^2\phi_x^2.
\end{aligned}\eqno(5.2)$$

To apply the Hirota bilinear method for constructing multi-periodic
wave solutions of the equation (5.2),   we  add  two variables and
consider a  general variable transformation
$$\begin{aligned}
&\phi=u+\theta_2v,\ \ u=i\partial_x\ln\frac{f}{g}, \ \
v=v_0-\partial_x\mathfrak{D}\ln(fg),
\end{aligned}\eqno(5.3)$$
where $u(x,t,\theta_1), f(x,t,\theta_1),
g(x,t,\theta_1):\mathbb{R}_{\Lambda}^{2,1}\rightarrow
\mathbb{R}_{\Lambda}^{1,0}$, and $v(x,t,\theta_1),
v_0=v_0(\theta_1):\mathbb{R}_{\Lambda}^{2,1}\rightarrow
\mathbb{R}_{\Lambda}^{0,1}$ is a special  solution of the equation
(5.2). Hereafter we use $\mathfrak{D}=\mathfrak{D}_1$ for
simplicity, Substituting (5.3) into (5.2), we then get the following
bilinear form
$$\begin{aligned}
&F( D_t, D_x)f\cdot
g=(D_t+D_x^3) f\cdot g=0,\\
&G(S_x, D_t, D_x)f\cdot g=(S_xD_t+S_xD_x^3+3v_0D_x^2+c) f\cdot g=0,
\end{aligned}\eqno(5.4)$$ where  $c=c(\theta_1,t):\mathbb{R}_{\Lambda}^{2,1}\rightarrow
\mathbb{R}_{\Lambda}^{0,1}$ is an odd integration constant to
variable $x$;  The equation (5.4) is
 a  type of coupled bilinear equations, which is more difficult to be dealt with than
 the bilinear equation (4.3) due to appearance of two functions and two equations. We will take full  advantages of
 Theorem 2 to reduce the number of constraint  equations.

Now we take into account  the periodicity of the solution (5.3), in
which we take
 $f$  and $g$   as
 $$f=\vartheta(\boldsymbol{\xi}+\boldsymbol{e}, \boldsymbol{\tau}),
 \ \ g=\vartheta(\boldsymbol{\xi}+\boldsymbol{h}, \boldsymbol{\tau}),
 \ \ \boldsymbol{e},\boldsymbol{h}\in \mathbb{Z}^N, $$
where  phase variable $\xi$  is taken as the form $\boldsymbol{\xi}
=(\xi_1,\cdots, \xi_N)^T$, $ \xi_j=\alpha_jx+\omega_j
t+\beta_j\theta_1\sigma +\delta_j, \ \ j=1,2, \cdots, N.$  By means
of Proposition 3, we deduce that
$$\begin{aligned}
&\phi=u+\theta_2v,\ \ u=i\sum_{k=1}^N\alpha_k
\partial_{\xi_k}\ln\frac{\vartheta(\boldsymbol{\xi}+\boldsymbol{e}, \boldsymbol{\tau})}
{\vartheta(\boldsymbol{\xi}+\boldsymbol{h}, \boldsymbol{\tau})},\\
&v=v_0-2\sum_{k,l=1}^N\alpha_k(\beta_l\sigma+\theta\alpha_l)
\partial_{\xi_k\xi_l}^2\ln[{\vartheta(\boldsymbol{\xi}+\boldsymbol{e}, \boldsymbol{\tau})}
{\vartheta(\boldsymbol{\xi}+\boldsymbol{h}, \boldsymbol{\tau})}],
\end{aligned}$$
which indicates that  the solution  $\phi$ is a quasi-periodic
function with $2N$ fundamental periods $\{\boldsymbol{e_j},\ \ j=1,
\cdots, N\}$ and $\{i\boldsymbol{\tau_j}, \ \ j=1, \cdots, N\}$.

 In the special case of $v_0=c=0$, starting
from the bilinear equation (5.4),  Zhang et al. found that the
equation (5.2) admits one-soliton solution \cite{Zhang}
$$\phi_1=i\partial_x\ln\frac{f_1}{g_1}+\theta_2[v_0-\partial_x\mathfrak{D}\ln (f_1g_1)],\eqno(5.5)$$
with
$$f_1=1+e^{\eta}, \ \ g_1=1-e^{\eta}$$
and  phase variable $\eta= p x-p^3 t+q\theta_1 \sigma+r$ with $p, q,
r\in\Lambda_0$. While two-soliton solution takes the form
$$\phi_1=i\partial_x\ln\frac{f_2}{g_2}+\theta_2[v_0-\partial_x\mathfrak{D}\ln(f_2g_2)],\eqno(5.6)$$
with
$$\begin{aligned}
&f_2=1+e^{\eta_1}+e^{\eta_2}+e^{\eta_1+\eta_2+A_{12}},\\
&g_2=1-+e^{\eta_1}-e^{\eta_2}+e^{\eta_1+\eta_2+A_{12}},
\end{aligned}\eqno(5.7)$$
and $\eta_j=p_jx-p_j^3t+q_j\theta_1 \sigma+r_j, \ \ j=1,2,$
\begin{eqnarray*}
&& e^{A_{12}}=\frac{(p_1-p_2)^2}{(p_1+p_2)^2
}+2\theta_1\sigma\frac{(p_1-p_2)(p_1q_2-p_2q_1)}{(p_1+p_2)^2 },
\end{eqnarray*}
here $p_j, q_j, r_j\in\Lambda_0, j=1, 2$ are free constants.\\[12pt]
%%%%%%%%%%%%%%%%%%%%%%%%%%%%%%%%%%%%%%%%%%%%%%%%%%%%%%%%%%%%%%%%%%%%%%%%%%%%%%%%%%%%%%%%%%%%%%%%%%%%%%%%%%%%%%%%%%%%%%%%%%%%%%%%%%%
{\bf 4.1. One-periodic waves and  asymptotic analysis}\\

 We first construct one-periodic wave solutions of the equation
 (5.2).  As a  simple case of  the theta function
 (3.2)  when $N=1, s=0$,  we take $f$ and $g$ as
$$\begin{aligned}
&f=\vartheta(\xi,0,0|\tau)=\sum_{n\in\mathbb{Z}}\exp({2\pi
in\xi-\pi n^2\tau}),\\
&g=\vartheta(\xi,1/2,0|\tau)=\sum_{n\in\mathbb{Z}}\exp({2\pi
in(\xi+1/2)-\pi n^2\tau})\\
&\ \ \ =\sum_{n\in\mathbb{Z}}(-1)^n\exp({2\pi in\xi-\pi
n^2\tau}),\end{aligned}\eqno(5.8)$$
 where the phase variable $\xi=\alpha
x+\omega t+\beta\theta_1\sigma+\delta$, and the parameter $\tau>0$.

Due to the fact that  $F(D_t, D_x)$ is an odd function, its
constraint equations in the formula (3.10)  vanish automatically for
$\mu=0$. Similarly the constraint equations associated with $G(S_x,
D_t, D_x)$ also vanish automatically for $\mu=1$. Therefore, the
Riemann theta function (5.8) is a solution of the bilinear equation
(5.4), provided the following equations
 $$\begin{aligned}
&\sum_{n\in\mathbb{Z}}\{2\pi i(2n-\mu)\omega-i(2\pi
\alpha)^3(2n-\mu)^3\}\exp[-2\pi\tau
(n-\mu/2)^2+\pi i(n-\mu/2)]|_{\mu=1}=0,\\
&\sum_{n\in\mathbb{Z}}\{-[2\pi(2n-\mu)]^2(\beta\sigma+\theta_1\alpha)\omega
+(2\pi (2n-\mu)]^4 (\beta\sigma+\theta_1\alpha)\alpha^3-(2\pi(2
n-\mu)\alpha]^2v_0+c\}\\
&\times\exp(-2\pi \tau (n-\mu/2)^2+\pi i(n-\mu/2))|_{\mu=0}=0.
\end{aligned}\eqno(5.9)$$

 We introduce the notations by
$$\begin{aligned}
&\lambda=e^{-\pi\tau/2 },\\
&\vartheta_1(\xi,\lambda)=\vartheta(2\mathbf{\xi},1/4,-1/2|
2\tau)=\sum_{n\in\mathbb{Z}}\lambda^{(2n-1)^2}\exp[4i\pi(n-1/2)
(\xi+1/4)]
,\\
&\vartheta_2(\xi,\lambda)=\vartheta(2\xi, 1/4,0 |
2\tau)=\sum_{n\in\mathbb{Z}}\lambda^{4n^2}\exp[4i\pi n(\xi+1/4)]
,\end{aligned}$$
 the equation (5.9) can be written as
a linear system about $\omega, c$
 $$\begin{aligned}
&\vartheta_1'\omega+\alpha^3\vartheta_1'''=0,\\
&(\beta\sigma+\theta_1\alpha)\vartheta_2''\omega+\vartheta_2c+(\beta\sigma+\theta_1\alpha)\alpha^3\vartheta_2^{(4)}
+\alpha^2\vartheta_2''v_0=0.
\end{aligned}\eqno(5.10)$$
where $\omega\in\Lambda_0$ is even and $c, v_0\in\Lambda_1$ are odd,
and we have denoted the derivative value of
$\vartheta_j(\xi,\lambda)$ at $\xi=0$ by simple notations
$$\vartheta_j'=\vartheta_j'(0,\lambda)=\frac{d\vartheta_j(\xi,\lambda)}{d\xi}|_{\xi=0}, \ \ j=1,2.$$
Moreover, we see that the functions $\vartheta_j$ and their
derivatives are independent of Grassmann variable $\theta$ and
anticommuting number $\sigma$.

We take $v_0=\gamma\alpha(\sigma+\theta\alpha), \ \gamma\in
\Lambda_0$ for the simplicity, then the system (5.10)  admits a
solution
 $$\begin{aligned}
&\omega=-\frac{\alpha^3\vartheta_1'''} {\vartheta_1'},\ \ \
c=\frac{\alpha^3(\beta\sigma+\theta_1\alpha)}{\vartheta_1'\vartheta_2}
(\vartheta_1'''\vartheta_2''-\vartheta_1'\vartheta_2^{(4)}-\gamma\vartheta_1'\vartheta_2''),
\end{aligned}\eqno(5.11)$$
where $\omega$ is independent of Grassmann variable $\theta$ and
auticommuting number $\sigma$. In this way,  a one-periodic wave
solution reads
$$\phi=i\partial_x\ln\frac{\vartheta(\xi,0,0|\tau)}{\vartheta(\xi,1/2,0|\tau)}
+\theta_2\{v_0-\partial_x\mathfrak{D}\ln
[\vartheta(\xi,0,0|\tau)\vartheta(\xi,1/2,0|\tau)]\},\eqno(5.12)$$
 where parameters $\omega$ and $c$ are given
by (5.11),  while  other parameters $\alpha, \beta, \tau,
\delta\in\Lambda_0$ are free. Among them, the three parameters
$\alpha$ and $\tau$ completely dominate a one-periodic wave.

In summary, one-periodic wave (5.12)  has the following features:

(i) It is one-dimensional and has two fundamental periods $1$ and
$i\tau$  in phase variable $\xi$. It can be viewed as a parallel
superposition of overlapping one-soliton waves, placed one period
apart (see Figure 5-7).

(ii) As in the case of the supersymmetric Sawada-Kotera-Ramani
equation, there is also  an influencing band among the real part of
one-periodic waves for the supersymmetric KdV equation under the
presence of the Grassmann variable (see Figure 4).

(iii) It was not observed  that influencing band appears among the
imaginary part and modulus of the one-periodic wave.  Moreover, they
seem to have the same shape from the observation of their plots (see
Figures 5 and 6).

\input epsf
\begin{figure}
\centering {\footnotesize $(a) \ \ \ \ \ \ \ \ \ \   \ \ \ \ \ \ \ \
\ \ \ \ \  \ \ \ \ \ \  $}\  \ \  \ \ \ \ \ \ \ \ \ \ \ \
 \ \ {\footnotesize $(b)$}$ \ \ \ \ \ \ \  \ \ \ \
 \ \ \ \ $\\
\includegraphics[width=2.1 in,angle=0]{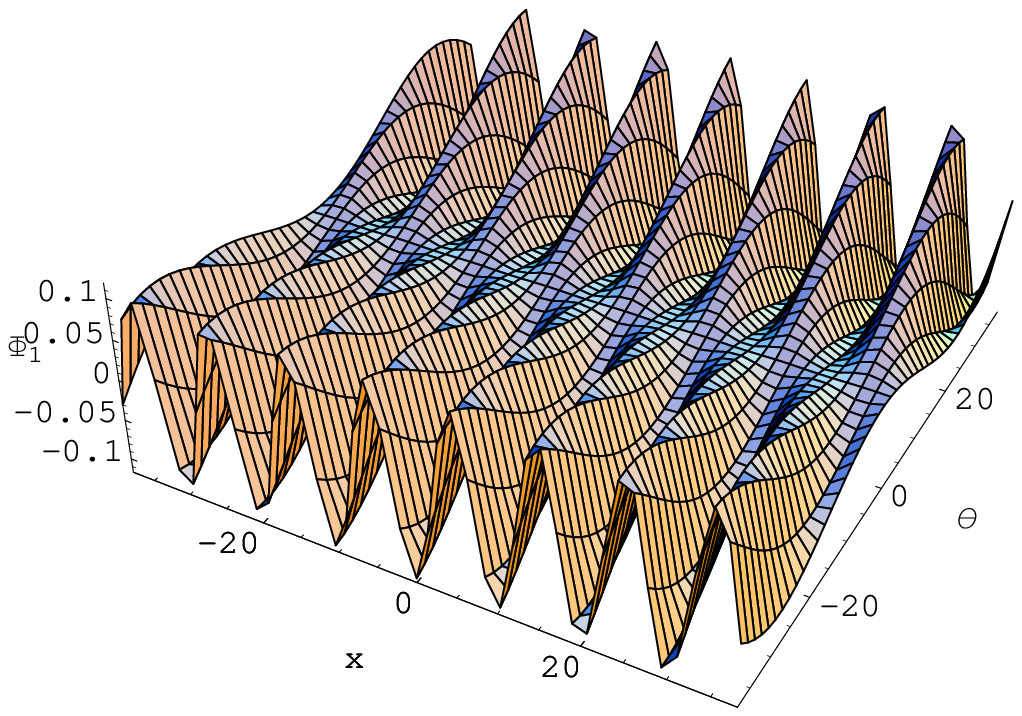}\ \ \ \ \ \  \ \ \ \
\ \
\includegraphics[width=1.4 in,angle=0]{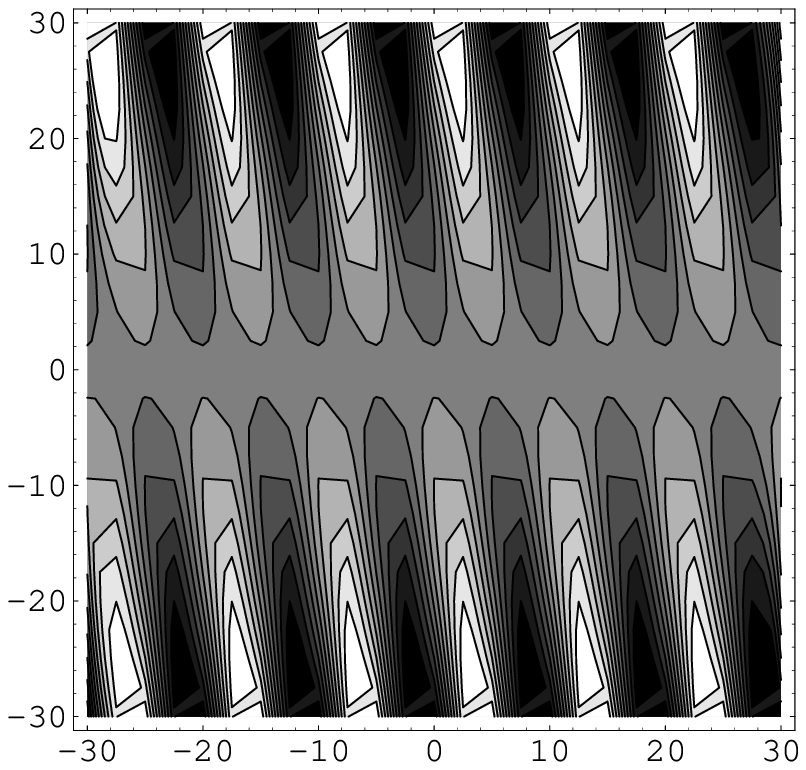}

\caption{\footnotesize\baselineskip=15pt  Real part of  one-periodic
wave for $\mathcal{N}=2$ supersymmetric KdV equation with
parameters: $\alpha=0.1,
 $ $\tau=3, \sigma_1=0.01$.  (a) Perspective view of wave. (b) Overhead view
of wave, with contour plot shown.  }
\end{figure}

\input epsf
\begin{figure}
\centering {\footnotesize $(a) \ \ \ \ \ \ \ \ \ \   \ \ \ \ \ \ \ \
\ \ \ \ \  \ \ \ \ \ \  $}\  \ \  \ \ \ \ \ \ \ \ \ \ \ \
 \ \ {\footnotesize $(b)$}$ \ \ \ \ \ \ \  \ \ \ \
 \ \ \ \ $\\
\includegraphics[width=2.1 in,angle=0]{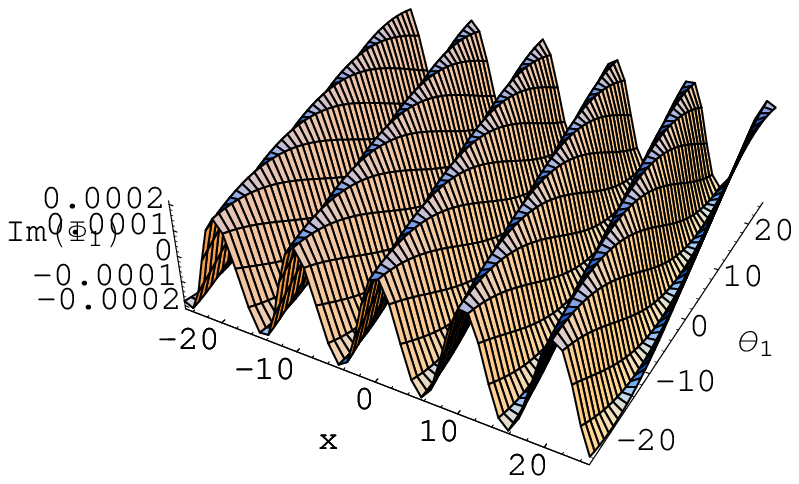}\ \ \ \ \ \  \ \ \ \
\ \
\includegraphics[width=1.4 in,angle=0]{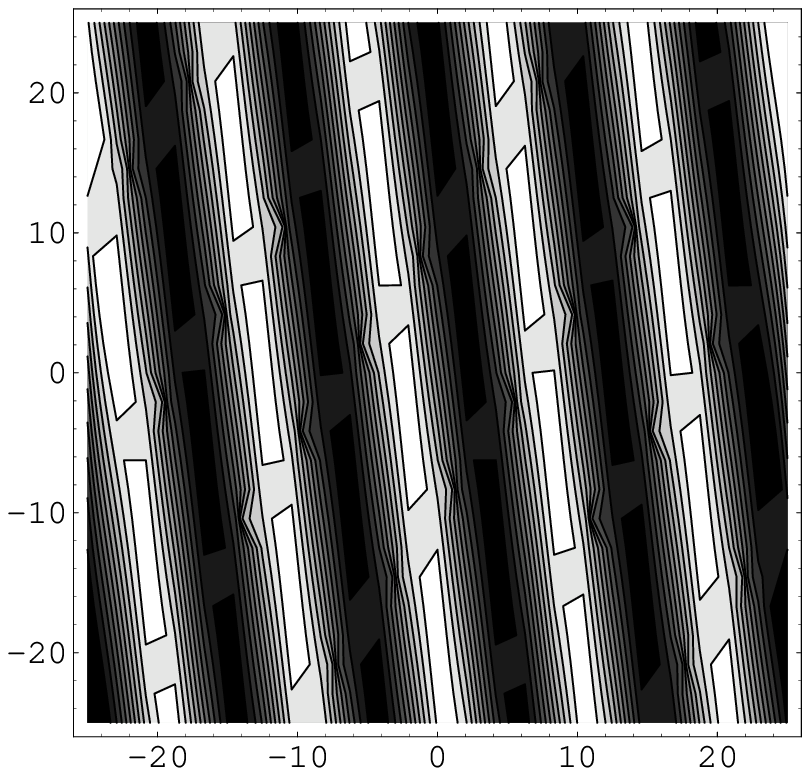}

\caption{\footnotesize\baselineskip=15pt Imaginary  part of
one-periodic wave for $\mathcal{N}=2$ supersymmetric KdV equation
with parameters: $\alpha=0.1,
 $ $\tau=3, \sigma_1=0.01$.   (a) Perspective view of wave. (b) Overhead view
of wave, with contour plot shown. }
\end{figure}

\input epsf
\begin{figure}
\centering {\footnotesize $(a) \ \ \ \ \ \ \ \ \ \   \ \ \ \ \ \ \ \
\ \ \ \ \  \ \ \ \ \ \  $}\  \ \  \ \ \ \ \ \ \ \ \ \ \ \
 \ \ {\footnotesize $(b)$}$ \ \ \ \ \ \ \  \ \ \ \
 \ \ \ \ $\\
\includegraphics[width=2.1 in,angle=0]{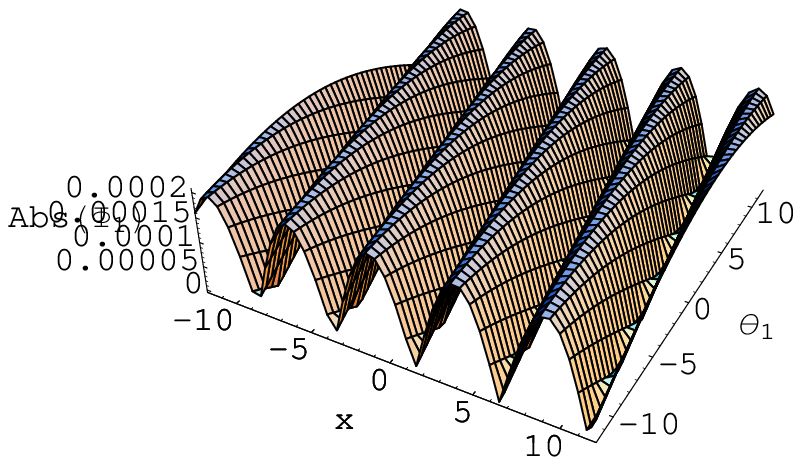}\ \ \ \ \ \  \ \ \ \
\ \
\includegraphics[width=1.4 in,angle=0]{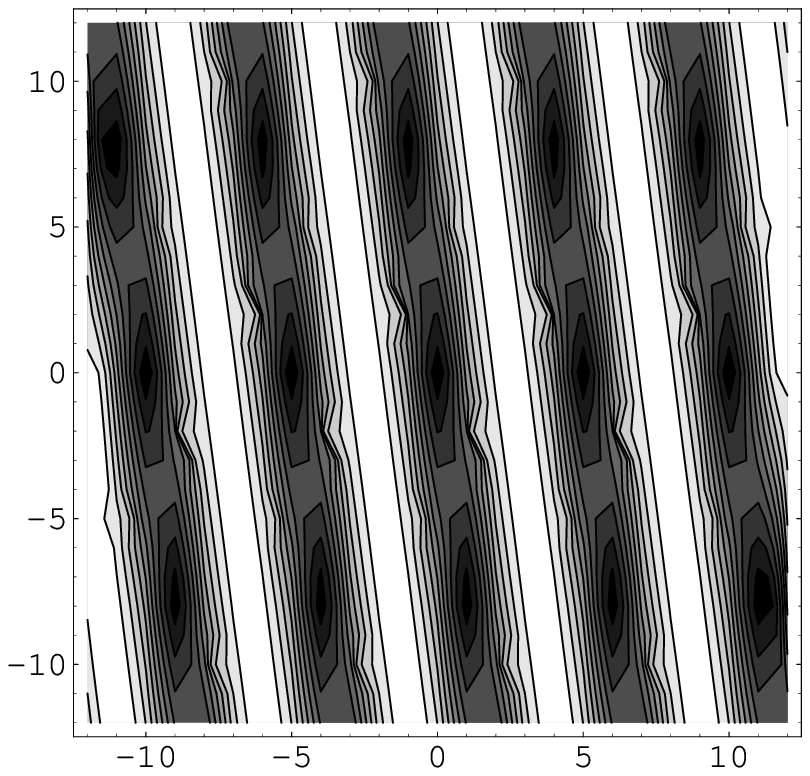}

\caption{\footnotesize\baselineskip=15pt  Modulus  of  one-periodic
wave for $\mathcal{N}=2$ supersymmetric KdV equation with
parameters: $\alpha=0.1,
 $ $\tau=3,\sigma_1=0.01$.   (a) Perspective view of wave. (b) Overhead view
of wave, with contour plot shown.}
\end{figure}

In the following, we further consider asymptotic properties of the
one-periodic wave solution. The relation between the one-periodic
wave solution (5.12) and the one-soliton solution (5.5) can be
established  as follows.

{\bf Theorem 5.}  Suppose that $\omega$ and $c$ are given by (5.11).
In the one-periodic wave solution (5.12), we choose parameters as
$$\gamma=0, \ \ \alpha=\frac{p}{2\pi i}, \ \ \beta=\frac{q}{2\pi i},\ \
\delta=\frac{r+\pi \tau}{2\pi i},\eqno(5.13)$$ where the $p, q$ and
$r$ are the same as those  in (5.5). Then we have the following
asymptotic properties
$$c\longrightarrow 0, \ \  \xi\longrightarrow\frac{\eta+\pi\tau}{2\pi i}, \ \
f\longrightarrow 1+e^{\eta}, \ \ g\longrightarrow 1-e^{\eta}, \ \
{\rm as } \ \ \lambda\rightarrow 0.$$   In other words, the
one-periodic solution (5.12) tends to the one-soliton solution (5.5)
 under a small amplitude limit , that is,
$$\phi\longrightarrow \phi_1, \ \ {\rm as } \ \
\lambda\rightarrow 0.\eqno(5.14)$$

{\it Proof.}  Here we will directly use the system (5.10) to analyze
asymptotic properties of one-periodic solution (5.12).  We
explicitly  expand the coefficients of system (5.10) as follows
$$\begin{aligned}
&\vartheta_1'=-4\pi\lambda+12\pi\lambda^9+\cdots,\quad
\vartheta_1'''=16\pi^3\lambda+432\pi^3\lambda^9+\cdots,
\\
&\vartheta_2=-2+2\lambda^4+\cdots,\quad
\vartheta_2''=32\pi^2\lambda^4+\cdots,\\
&\vartheta_2^{(4)}=512\pi^4\lambda^4+\cdots\end{aligned}\eqno(5.15)$$
Suppose that  the solution of the system (5.10) is of the form
$$\begin{aligned}
&\omega=\omega_0+\omega_1\lambda+\omega_2\lambda^2+\cdots=\omega_0+o(\lambda),\\
&c=c_0+c_1\lambda+c_2\lambda^2+\cdots=c_0+o(\lambda).
\end{aligned}\eqno(5.16)$$

Substituting the expansions (5.15) and (5.16) into the system (5.11)
and letting $\lambda\longrightarrow 0$, we immediately obtain
 the following relations
$$
 \begin{aligned}
 &-4\pi\omega_0+16\pi^3\alpha^3=0,\ \ c_0=0,
 \end{aligned}$$
which has a solution
 $$c_0=0, \ \ w_0=4\pi^2\alpha^3.\eqno(5.17)$$
Combining  (5.13) and (5.17) leads to
$$c\longrightarrow 0, \ \  2\pi i\omega\longrightarrow 8\pi^3i\alpha^3=-p^3, \ \ {\rm as } \ \
\lambda\rightarrow 0,$$ or equivalently
$$\begin{aligned}
&\hat{\xi}=2\pi i\xi-\pi \tau=p x+2\pi i\omega t+q\theta_1\sigma+r\\
&\quad \longrightarrow px-p^3t+q\theta_1\sigma+r=\eta,\ \ {\rm as}\
\ \lambda\rightarrow 0.
\end{aligned}\eqno(5.18)$$

 It remains to  identify   that the one-periodic wave  (5.12) possesses  the same
form with the one-soliton solution (5.5) under the limit
$\lambda\rightarrow 0$. For this purpose, we start to  expand the
 functions $f$  and $g$ in the form
$$ f=1+\lambda^2(e^{2\pi i\xi}+e^{-2\pi i\xi})+\lambda^8(e^{4\pi i\xi}+e^{-4\pi i\xi})
+\cdots .$$
$$ g=1-\lambda^2(e^{2\pi i\xi}+e^{-2\pi i\xi})+\lambda^8(e^{4\pi i\xi}+e^{-4\pi i\xi})
+\cdots .$$
 By using   (5.13) and (5.17),  it follows that
$$\begin{aligned}
&f=1+e^{\hat{\xi}}+\lambda^4(e^{-\hat{\xi}}+e^{2\hat{\xi}})+\lambda^{12}(e^{-2\hat{\xi}}+e^{3\hat{\xi}})
+\cdots\\
&\quad \longrightarrow 1+e^{\hat{\xi}}\longrightarrow 1+e^{\eta},\ \
{\rm as}\ \ \lambda\rightarrow 0;\\
&g=1-e^{\hat{\xi}}+\lambda^4(e^{2\hat{\xi}}-e^{-\hat{\xi}})+\lambda^{12}(e^{-2\hat{\xi}}-e^{3\hat{\xi}})
+\cdots\\
&\quad \longrightarrow 1-e^{\hat{\xi}}\longrightarrow 1-e^{\eta},\ \
{\rm as}\ \ \lambda\rightarrow 0.
\end{aligned}\eqno(5.19)$$
The expression (5.13) follows from  (5.19), and thus we conclude
that the one-periodic solution (5.12) just goes to the one-soliton
solution (5.5) as the amplitude
$\lambda\rightarrow 0$. $\square$\\[12pt]
%%%%%%%%%%%%%%%%%%%%%%%%%%%%%%%%%%%%%%%%%%%%%%%%%%%%%%%%%%%%%%%%%%%%%%%%%%%%%%%%%%%%%%%%%%%%%%%%%%%%%%%%%%%%%%%%%%%%%%%%%%%%%%%%%%%%%%
%%%%%%%%%%%%%%%%%%%%%%%%%%%%%%%%%%%%%%%%%%%%%%%%%%%%%%%%%%%%%%%%%%%%%%%%%%%%%%%%%%%%%%%%%%%%%%%%%%%%%%%%%%%%%%%%%%%%%%%%%%%%%%%%%%%%%%
{\bf 4.2. Two-periodic waves and  asymptotic properties}\\

We now  consider  two-periodic wave solutions to the supersymmetric
KdV equation (5.2).  For the case when $N=2, \
\boldsymbol{s}=\boldsymbol{0},\
\boldsymbol{\varepsilon}=\boldsymbol{1/2}=(1/2,1/2)$ in the Riemann
theta function (3.2), we choose  $f$ and $g$ to be
$$\begin{aligned}
&f=\vartheta(\boldsymbol{\xi},\boldsymbol{0},\boldsymbol{0}|
\boldsymbol{\tau})=\sum_{\boldsymbol{n}\in \mathbb{Z}^2} \exp\{2\pi
i\langle\boldsymbol{\xi},\boldsymbol{n}\rangle-\pi\langle\boldsymbol{\tau}\boldsymbol{ n},\boldsymbol{n}\rangle\},\\
 &g=\vartheta(\boldsymbol{\xi},\boldsymbol{1/2},\boldsymbol{0}|
\boldsymbol{\tau})=\sum_{\boldsymbol{n}\in \mathbb{Z}^2} \exp\{2\pi
i\langle\boldsymbol{\xi}+\boldsymbol{1/2},\boldsymbol{n}\rangle-\pi\langle\boldsymbol{\tau}\boldsymbol{
n},\boldsymbol{n}\rangle\}\\
&\ \ =\sum_{\boldsymbol{n}\in \mathbb{Z}^2}(-1)^{n_1+n_2} \exp\{2\pi
i\langle\boldsymbol{\xi},\boldsymbol{n}\rangle-\pi\langle\boldsymbol{\tau}\boldsymbol{
n},\boldsymbol{n}\rangle\}
\end{aligned}\eqno(5.20)$$
where we denote  $\boldsymbol{n}=(n_1, n_2)\in Z^2,\ \
\boldsymbol{\xi}=(\xi_1, \xi_2)\in\mathcal{C}^2, \ \
\xi_i=\alpha_jx+\omega_jt+\beta_j\theta_1\sigma+\delta_j, \ \ j=1,
2$, and $ \boldsymbol{\alpha}=(\alpha_1, \alpha_2),\
\boldsymbol{\omega}=(\omega_1, \omega_2),\
\boldsymbol{\beta}=(\beta_1, \beta_2) \in \mathcal{C}^2,$; The
matrix  $\boldsymbol{\tau}$ is a  positive definite  and real-valued
symmetric ${2\times 2}$ matrix, which can  take  the form
$$\boldsymbol{\tau}=(\tau_{ij})_{2\times 2}, \ \ \tau_{12}=\tau_{21},\ \ \tau_{11}>0,\ \
 \tau_{22}>0, \ \ \tau_{11}\tau_{22}-\tau_{12}^2>0.$$

 According to Theorem 5, constraint equations associated with  $F(D_t, D_x)$
 vanish automatically for $(\mu_1, \mu_2)=(0,0),
(1, 1)$,  and the constraint equations associated with $G(S_x, D_t,
D_x)$ vanish automatically for $(\mu_1, \mu_2)=(1,0), (0, 1)$.
Hence, making the theta functions $f$ and $g$ satisfy the bilinear
equation (5.4) gives to  the following  constraint equations
$$\begin{aligned}
&\sum_{n_1,n_2\in \mathbb{Z}}\left[2\pi i\langle
2\boldsymbol{n}-\boldsymbol{\mu},\boldsymbol{\omega}\rangle-8\pi^3i\langle
2\boldsymbol{n}-\boldsymbol{\mu},
\boldsymbol{\alpha}\rangle^3\right]\exp\{-2\pi \langle
\boldsymbol{\tau} (\boldsymbol{n}-\boldsymbol{\mu}/2),
\boldsymbol{n}-\boldsymbol{\mu}/2\rangle\\
&+\pi
i\sum_{j=1}^{2}(n_j-\mu_j/2)\}|_{\boldsymbol{\mu}=(\mu_1,\mu_2)}=0,
\ \ {\rm for}\ \ (\mu_1,\mu_2)=(0,1),\ (1,0).
\end{aligned}\eqno(5.21)$$
and
  $$\begin{aligned}
&\sum_{n_1,n_2\in \mathbb{Z}}[-4\pi^2\langle
2\boldsymbol{n}-\boldsymbol{\mu},\sigma\boldsymbol{\beta}+\theta_1\boldsymbol{\alpha}\rangle
\langle
2\boldsymbol{n}-\boldsymbol{\mu},\boldsymbol{\omega}\rangle+16\pi^4\langle
2\boldsymbol{n}-\boldsymbol{\mu},\sigma\boldsymbol{\beta}+\theta_1\boldsymbol{\alpha}\rangle
\langle
2\boldsymbol{n}-\boldsymbol{\mu},\boldsymbol{\alpha}\rangle^3\\
&\ \ \ \ -4\pi^2\langle
2\boldsymbol{n}-\boldsymbol{\mu},\boldsymbol{\alpha}\rangle^2v_0+c]
\exp\{-2\pi \langle \boldsymbol{\tau} (\boldsymbol{n}-\boldsymbol{\mu}/2), \boldsymbol{n}-\boldsymbol{\mu}/2\rangle\\
&\ \ \ \ +\pi
i\sum_{j=1}^{2}(n_j-\mu_j/2)\}|_{\boldsymbol{\mu}=(\mu_1,\mu_2)}=0,
\ \ {\rm for}\ \ (\mu_1,\mu_2)=(0,0),\ (1,1).
\end{aligned}\eqno(5.22)$$

 Next, introducing the following notations
 \begin{eqnarray*}
 &&\lambda_{kl}=e^{-\pi\tau_{kl}/2}, k, l=1,2, \boldsymbol{\lambda}=(\lambda_{11},\lambda_{12},
\lambda_{22})\\
&&\vartheta_j(\boldsymbol{\xi},\boldsymbol{\lambda})=\vartheta(2\xi,\boldsymbol{1/4},-\boldsymbol{s}_j/2|2\tau)
 =\sum_{n_1,n_2\in Z}\exp\{4\pi i\langle\boldsymbol{\xi}+\boldsymbol{1/4} ,\boldsymbol{n}-\boldsymbol{s_j}/2\rangle\}
 \prod_{k,l=1}^{2}\lambda_{kl}^{(2n_k-s_{j,k})(2n_j-s_{j,l})} ,\\
&& \boldsymbol{s_j}=(s_{j,1}, s_{j,2}),\quad j=1, 2, 3, 4,\ \
\boldsymbol{s_1}=(0,1),\ \ \boldsymbol{s_2}=(1,0),\ \
\boldsymbol{s_3}=(0,0),\ \ \boldsymbol{s_4}=(1,1),
\end{eqnarray*}
then by using (3.15), the  system (5.21) and (5.22) can be written
as a linear system
 $$\begin{aligned}
&
(\boldsymbol{\omega}\cdot\nabla)\vartheta_j(0,\boldsymbol{\lambda})+
 (\boldsymbol{\alpha}\cdot\nabla)^3\vartheta_j(0,\boldsymbol{\lambda})=0,\
 j=1, 2,\\
 &[(\sigma\boldsymbol{\beta}+\theta_1\boldsymbol{\alpha})\cdot\nabla]
 (\boldsymbol{\omega}\cdot\nabla)\vartheta_j(0,\boldsymbol{\lambda})+
 [(\sigma\boldsymbol{\beta}+\theta_1\boldsymbol{\alpha})\cdot\nabla]
 (\boldsymbol{\alpha}\cdot\nabla)^3\vartheta_j(0,\boldsymbol{\lambda})\\
 &+(\boldsymbol{\alpha}\cdot\nabla)^2\vartheta_j(0,\boldsymbol{\lambda})v_0
 +\vartheta_j(0,\boldsymbol{\lambda})c=0, \ j=3,4.
 \end{aligned}\eqno(5.23)$$
 This system can be solved in such a way: After we obtain $\omega,_1, \omega_2$ form the first
 two equations, We substitute them into last two equations to  get
 $v_0, c$. With the solution $(\omega_1, \omega_2, v_0,
c)$, we get  a two-periodic wave solution to the supersymmetric
equation (5.2)
$$\phi=i\partial_x\ln\frac{\vartheta(\boldsymbol{\xi},\boldsymbol{0},\boldsymbol{0}|\boldsymbol{\tau})}
{\vartheta(\boldsymbol{\xi},\boldsymbol{1/2},\boldsymbol{0}|\boldsymbol{\tau})}
+\theta_2\{v_0-\partial_x\mathfrak{D}\ln
[\vartheta(\boldsymbol{\xi},\boldsymbol{0},\boldsymbol{0}|\boldsymbol{\tau})
\vartheta(\boldsymbol{\xi},\boldsymbol{1/2},\boldsymbol{0}|\boldsymbol{\tau})]\},\eqno(5.24)$$
where parameters $\omega_1, \omega_2, v_0$ and $ c$ are given by
(5.22), while other parameters $\sigma_{1}, \sigma_{2}, \alpha_1,$ $
\alpha_2, \tau_{11}$, $\tau_{22}, \tau_{12}, \delta_1$ and
$\delta_2$ are free.

In summary, the two-periodic wave (5.24), which is  a direct
generalization of one-periodic waves,  has the following features:

(i) Its surface pattern is two-dimensional, namely,  there are two
phase variables $\xi_1$ and $\xi_2$.   It has $4$ fundamental
periods $\{e_1, e_2\}$ and $\{i\tau_1, i\tau_2\}$ in $(\xi_1,
\xi_2)$, and  is spatially periodic in  two directions $\xi_1,
\xi_2$.  Its real part is not periodic in $\theta$ direction, while
its real part, imaginary part and modulus are all periodic in
   $x$ and   $t$   directions.

(iii)  There is also  an influencing band among the Real part of
two-periodic waves for the supersymmetric KdV equation under the
presence of the Grassmann variable ( see Figure 7 ).

(iv) It was not found that influencing band appears among the
imaginary part and modulus of two-periodic waves to the
supersymmetric KdV equation ( see Figures 8 and 9 ).

\input epsf
\begin{figure}
\centering {\footnotesize $(a) \ \ \ \ \ \ \ \ \ \   \ \ \ \ \ \ \ \
\ \ \ \ \  \ \ \ \ \ \  $}\  \ \  \ \ \ \ \ \ \ \ \ \ \ \
 \ \ {\footnotesize $(b)$}$ \ \ \ \ \ \ \  \ \ \ \
 \ \ \ \ $\\
\includegraphics[width=2.1 in,angle=0]{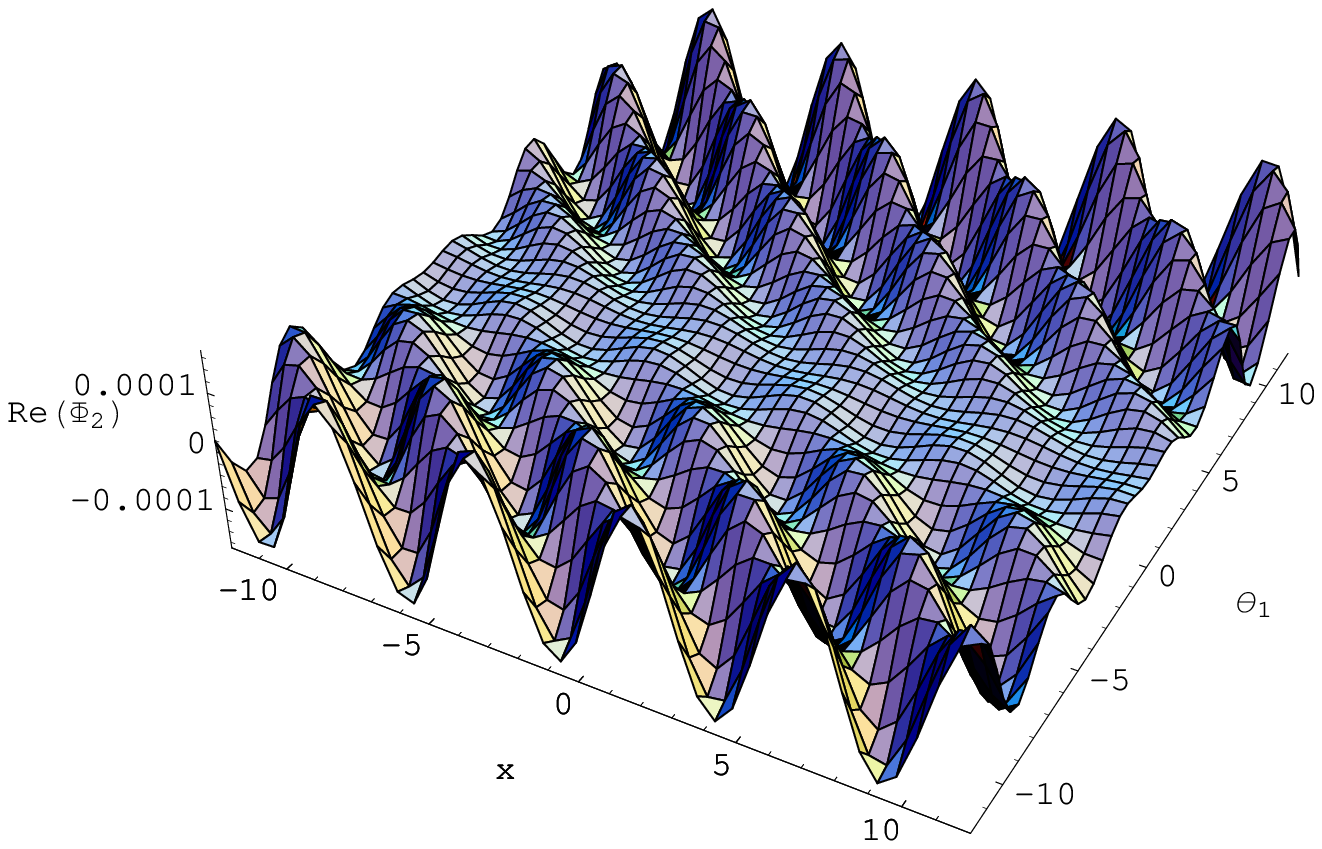}\ \ \ \ \ \  \ \ \ \
\ \
\includegraphics[width=1.4 in,angle=0]{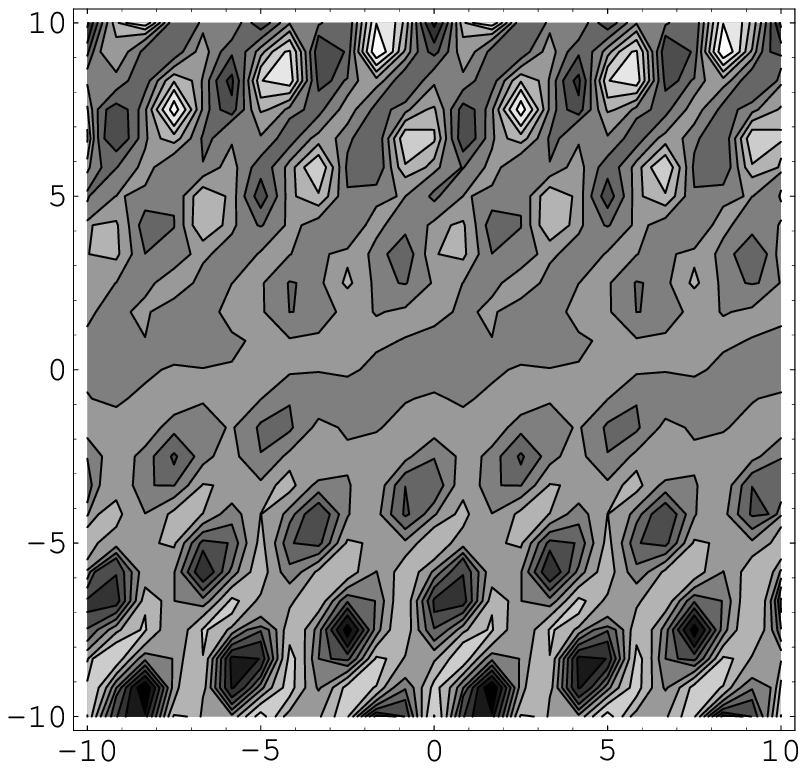}

\caption{\footnotesize\baselineskip=15pt  Real part of  two-periodic
wave for $\mathcal{N}=2$ supersymmetric KdV equation with
parameters: $\alpha_1=0.1,\ \alpha_2=0.2 $ $\tau_{11}=3, \
\tau_{12}=0.2, \tau_{22}=3, \sigma_1=0.2, \ \sigma_2=-0.1$. (a)
Perspective view of wave. (b) Overhead view of wave, with contour
plot shown.   }
\end{figure}

\input epsf
\begin{figure}
\centering {\footnotesize $(a) \ \ \ \ \ \ \ \ \ \   \ \ \ \ \ \ \ \
\ \ \ \ \  \ \ \ \ \ \  $}\  \ \  \ \ \ \ \ \ \ \ \ \ \ \
 \ \ {\footnotesize $(b)$}$ \ \ \ \ \ \ \  \ \ \ \
 \ \ \ \ $\\
\includegraphics[width=2.1 in,angle=0]{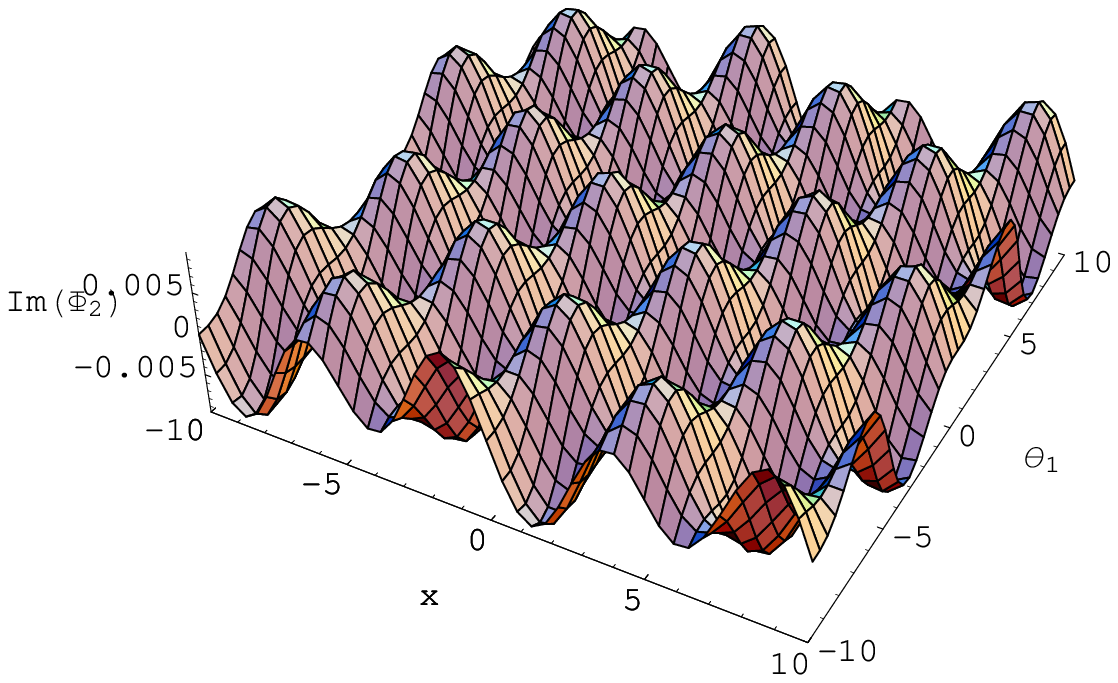}\ \ \ \ \ \  \ \ \ \
\ \
\includegraphics[width=1.4 in,angle=0]{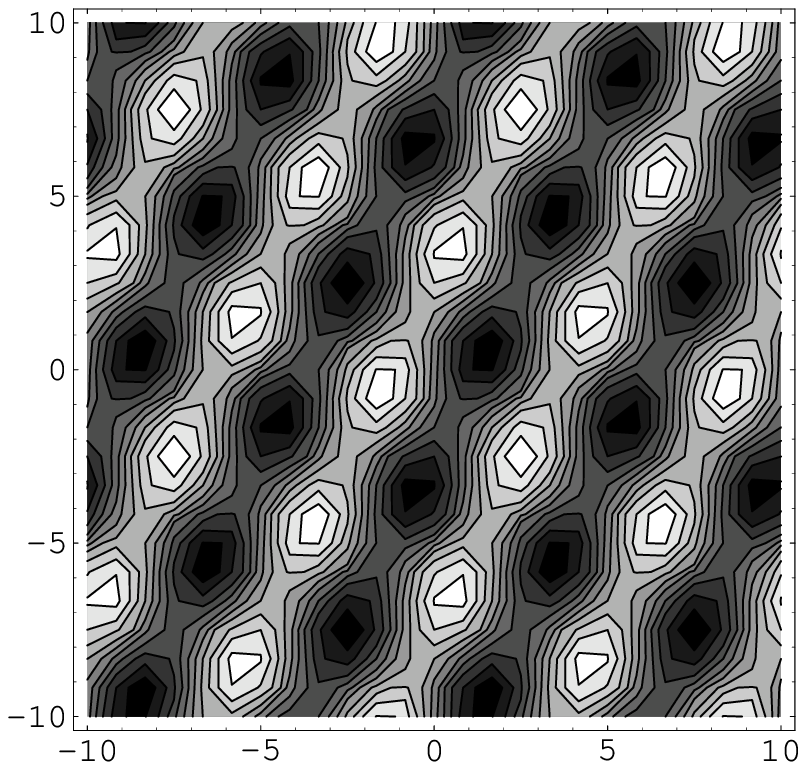}

\caption{\footnotesize\baselineskip=15pt Imaginary  part of
two-periodic wave for $\mathcal{N}=2$ supersymmetric KdV equation
with parameters: $\alpha_1=0.1,\ \alpha_2=0.2 $ $\tau_{11}=3, \
\tau_{12}=0.2, \tau_{22}=3, \sigma_1=0.2, \ \sigma_2=-0.1$.   (a)
Perspective view of wave. (b) Overhead view of wave, with contour
plot shown.  }
\end{figure}

\input epsf
\begin{figure}
\centering {\footnotesize $(a) \ \ \ \ \ \ \ \ \ \   \ \ \ \ \ \ \ \
\ \ \ \ \  \ \ \ \ \ \  $}\  \ \  \ \ \ \ \ \ \ \ \ \ \ \
 \ \ {\footnotesize $(b)$}$ \ \ \ \ \ \ \  \ \ \ \
 \ \ \ \ $\\
\includegraphics[width=2.1 in,angle=0]{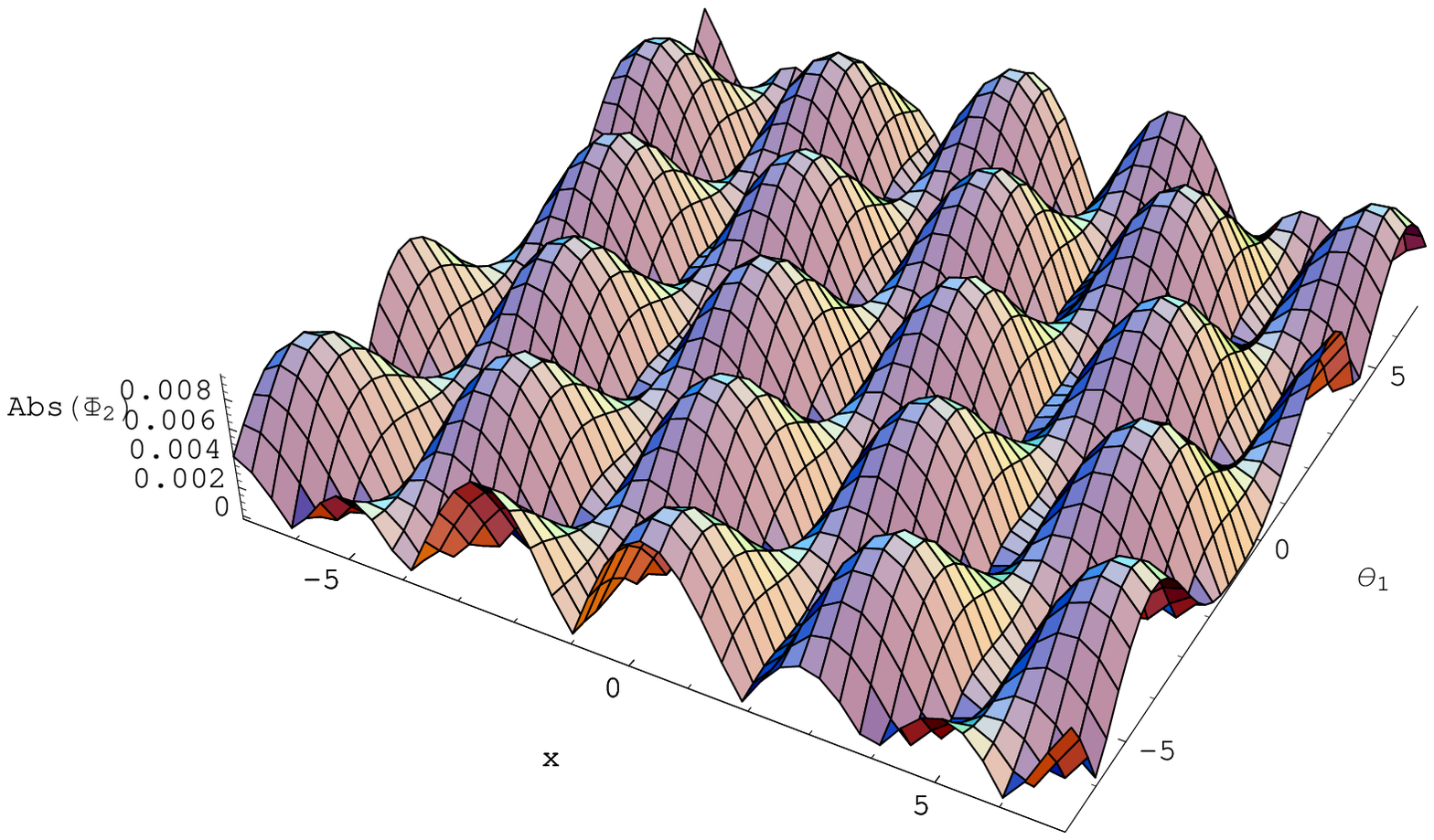}\ \ \ \ \ \  \ \ \ \
\ \
\includegraphics[width=1.4 in,angle=0]{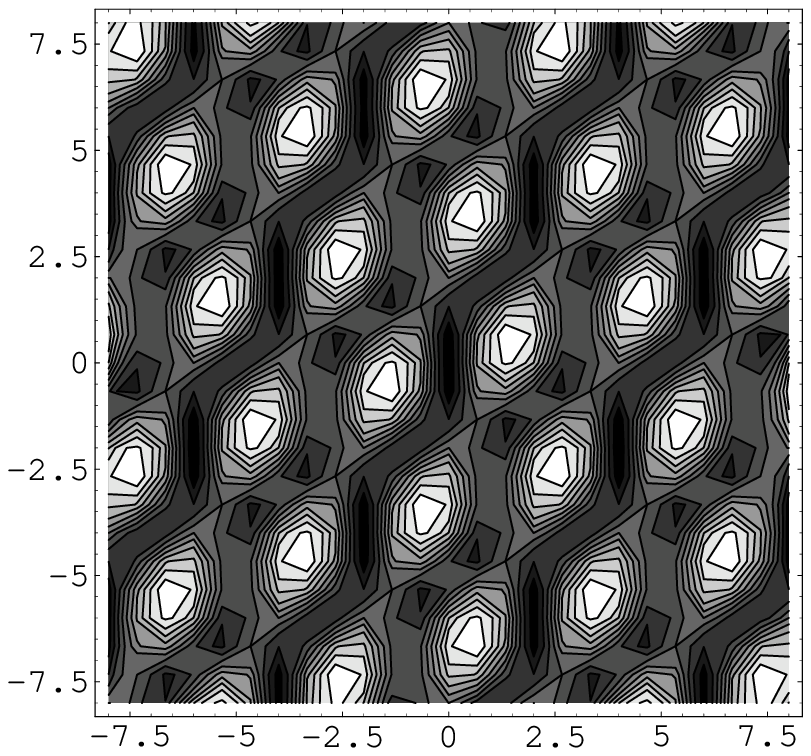}

\caption{\footnotesize\baselineskip=15pt  Modulus  of  two-periodic
wave for $\mathcal{N}=2$ supersymmetric KdV equation with
parameters: $\alpha_1=0.1,\ \alpha_2=0.2 $ $\tau_{11}=3, \
\tau_{12}=0.2, \tau_{22}=3, \sigma_1=0.2, \ \sigma_2=-0.1$.   (a)
Perspective view of wave. (b) Overhead view of wave, with contour
plot shown.   }
\end{figure}

Finally,  we consider the asymptotic properties of the two-periodic
solution (5.24).
 In a similar way to Theorem 5, we can
establish the relation between the two-periodic solution (5.24) and
the two-soliton solution (5.6) as follows.

{\bf Theorem 6.}   Assume that $(\omega_1, \omega_2, v_0, c)$ is a
solution of the system (5.23). In the two-periodic wave solution
(5.24),  a choice of parameters is given by
$$
\begin{aligned}
&\alpha_j=\frac{p_j}{2\pi i},\ \ \beta_j=\frac{q_j}{2\pi i}, \ \
\delta_j=\frac{r_j+\pi\tau_{jj} }{2\pi i},\ \
\tau_{12}=-\frac{A_{12}}{2\pi},\ \ j=1,2,
\end{aligned}\eqno(5.25)$$
 with the $p_j, q_j, r_j\in\Lambda_0,
j=1, 2$ and $A_{12}$  as those given in (5.6). Then under the
constraint $\alpha_1\beta_2=\alpha_2\beta_1$,  we have the following
asymptotic relations
$$
\begin{aligned}
&v_0\longrightarrow 0, \ \ \ c\longrightarrow 0, \ \
\xi_j\longrightarrow\frac{\eta_j+\pi \tau_{jj}}{2\pi i}, \ \
j=1, 2,\\
& f\longrightarrow 1+e^{\eta_1}+e^{\eta_2}+e^{\eta_1+\eta_2+A_{12}},
\  \ g\longrightarrow 1-e^{\eta_1}-e^{\eta_2}+e^{\eta_1+\eta_2+A_{12}},\\
& \  \ \ {\rm as} \ \ \lambda_{11}, \lambda_{22}\rightarrow 0.
\end{aligned}\eqno(5.26)$$
So the two-periodic wave  solution (5.24) just tends to the
two-soliton solution (5.6) under a certain limit
$$\phi\longrightarrow \phi_2, \ \ {\rm as }\ \ \lambda_{11}, \lambda_{22}\rightarrow 0.$$

{\it Proof.}   From (5.25), the constraint
$\alpha_1\beta_2=\alpha_2\beta_1$ leads to  $p_1q_2-p_2q_1=0$, which
implies that $\tau_{12}=-A_{12}/2\pi$ is independent of Grassmann
variable $\theta$ according to (5.7).

  Using (5.20), we explicitly expand  the  functions $f$  and $g$ in the following form
\begin{eqnarray*}
&& f=1+(e^{2\pi i\xi_1}+e^{-2\pi i\xi_1})e^{-\pi \tau_{11}}
+(e^{2\pi i\xi_2}+e^{-2\pi i\xi_2})e^{-\pi
\tau_{22}}\\
&&\ \ \ \ \ \ +(e^{2\pi i(\xi_1+\xi_2)}+e^{-2\pi
i(\xi_1+\xi_2)})e^{-\pi
(\tau_{11}+2\tau_{12}+\tau_{22})}+\cdots\\
&& g=1-(e^{2\pi i\xi_1}+e^{-2\pi i\xi_1})e^{-\pi \tau_{11}}
-(e^{2\pi i\xi_2}+e^{-2\pi i\xi_2})e^{-\pi
\tau_{22}}\\
&&\ \ \ \ \ \ +(e^{2\pi i(\xi_1+\xi_2)}+e^{-2\pi
i(\xi_1+\xi_2)})e^{-\pi (\tau_{11}+2\tau_{12}+\tau_{22})}+\cdots
\end{eqnarray*}
Further  using (4.5) and making a transformation
$\hat{\omega}_j=2\pi i \omega_j, j=1, 2$, we infer that
\begin{eqnarray*}
&&f=1+e^{\hat{\xi}_1}+e^{\hat{\xi}_2}+e^{\hat{\xi}_1+\hat{\xi}_2-2\pi
\tau_{12} }+\lambda_{11}^4e^{-\hat{\xi}_1}
 +\lambda_{22}^4e^{-\hat{\xi}_2}+\lambda_{11}^4\lambda_{22}^4e^{-\hat{\xi}_1-\hat{\xi}_2-2\pi \tau_{12}}+\cdots\\
&& \ \ \ \ \ \longrightarrow
1+e^{\hat{\xi}_1}+e^{\hat{\xi}_2}+e^{\hat{\xi}_1+\hat{\xi}_2+A_{12}},\
\ {\rm as}\ \ \lambda_{11}, \lambda_{22} \rightarrow 0,\\
&&g=1-e^{\hat{\xi}_1}-e^{\hat{\xi}_2}+e^{\hat{\xi}_1+\hat{\xi}_2-2\pi
\tau_{12} }-\lambda_{11}^4e^{-\hat{\xi}_1}
 -\lambda_{22}^4e^{-\hat{\xi}_2}+\lambda_{11}^4\lambda_{22}^4e^{-\hat{\xi}_1-\hat{\xi}_2-2\pi \tau_{12}}+\cdots\\
&& \ \ \ \ \ \longrightarrow
1-e^{\hat{\xi}_1}-e^{\hat{\xi}_2}+e^{\hat{\xi}_1+\hat{\xi}_2+A_{12}},\
\ {\rm as}\ \ \lambda_{11}, \lambda_{22} \rightarrow 0,
 \end{eqnarray*}
where $\hat{\xi}_j=p_jx+\hat{\omega}_jt+q_j\theta_1\sigma+r_j, \ \
j=1,2.$

 It remains to prove that
$$
\begin{aligned}
&c\longrightarrow 0, \ \ \ \hat{\omega}_j\longrightarrow-p_j^3, \ \
\hat{\xi}_j\longrightarrow \eta_j,\ \ j=1, 2, \ \ \ {\rm as} \ \
\lambda_{11}, \lambda_{22}\rightarrow 0.\end{aligned}\eqno(5.27)$$

As  in the case when $N=1$,  we let  the solution of the system
(5.23) be the form
$$\begin{aligned}
&\omega_1=\omega_{1,0}+\omega_{1,1}\lambda_{11}+\omega_{2,2}\lambda_{22}+o(\lambda_{11},\lambda_{22}),\\
&\omega_2=\omega_{2,0}+\omega_{2,1}\lambda_{11}+\omega_{2,2}\lambda_{22}+o(\lambda_{11},\lambda_{22}),\\
&v_0=v_{0,0}+v_{0,1}\lambda_{11}+v_{0,2}\lambda_{22}+o(\lambda_{11},\lambda_{22}),\\
&c=c_0+c_1\lambda_{11}+c_2\lambda_{22}+o(\lambda_{11},\lambda_{22}).
\end{aligned}\eqno(5.28)$$
Expanding  functions  $\vartheta_{j}, j=1,2,3,4$ in the system
(5.23), together with substitution of assumption (5.28), the second
and third equation is divided by $\lambda_{11}$ and $\lambda_{22}$,
respectively; the fourth equation is divided by
$\lambda_{11}\lambda_{22}$, and letting $v_{0,0}=0, \
\lambda_{11},\lambda_{22}\longrightarrow 0$ , we then obtain
$$\begin{aligned}
&c_{0}=0, \\
&-8\pi\omega_{1,0}+32\pi^4\alpha_1^3=0,\\
&-8\pi\omega_{2,0}+32\pi^4\alpha_2^3=0,\\
&[-8\pi^2(\omega_{1,0}+\omega_{2,0})+32\pi^4(\alpha_1+
\alpha_2)^3]\lambda_{12}\\
&\ \ \ \ +[-8\pi^2(\omega_{1,0}-\omega_{2,0})+32\pi^4(\alpha_1-
\alpha_2)^3]\lambda_{12}^{-1}=0.
\end{aligned}\eqno(5.29)$$
The first three equations in the system (5.9) have a  solution
$$
 \begin{aligned}
 &c_0=0, \ \ v_{0,0}=0, \ \ \omega_{1,0}=4\pi^2\alpha_1^3, \ \ \omega_{2,0}=4\pi^2\alpha_2^3,
 \end{aligned}\eqno(5.30)$$
The  fourth equation in the system (5.29) satisfied automatically by
using (5.25) and (5.30), thus we have
$$
 \begin{aligned}
 &c_0=c_1=c_2=0, \ \ v_{0,0}=0, \ \ \omega_{1,0}=4\pi^2\alpha_1^3, \ \
 \omega_{2,0}=4\pi^2\alpha_2^3.
 \end{aligned}\eqno(5.31)$$
Using (5.28) and (5.31),  we conclude  that
$$\begin{aligned}
&v_0=o(\lambda_{11}, \lambda_{22})\longrightarrow 0, \ \
c=o(\lambda_{11}, \lambda_{22})\longrightarrow 0,\ \
\omega_1=4\pi^2\alpha_1^3+o(\lambda_{11},
\lambda_{22})\longrightarrow
4\pi^2\alpha_1^3,\\
 &\omega_2=4\pi^2\alpha_2^3+o(\lambda_{11}, \lambda_{22})\longrightarrow
4\pi^2\alpha_2^3,  \ \ \ {\rm as } \ \ \lambda_{11},
\lambda_{22}\rightarrow 0,
\end{aligned}$$
and therefore  we have  (5.26).  So  the two-periodic wave solution
(5.23) tends to the two-supersoliton solution (5.6) as
$\lambda_{11}, \lambda_{22}\rightarrow 0$. $\square$\\ [12pt]
 %%%%%%%%%%%%%%%%%%%%%%%%%%%%%%%%%%%%%%%%%%%%%%%%%%%%%%%%%%%%%%%%%%%%%%%%%%%%%%%%%%%%
{\bf\large  6.  Conclusion and future work  }\\

Following the procedure described in this paper, we are able to
construct quasi-periodic wave solutions for other supersymmetric
equations also can be dealt with by the same way. For instance,

(1)\ Supersymmetric KdV equation \cite{Manin, Oevel, Carstea}
$$
 \begin{aligned}
 &\Phi_t+3\left(\Phi\mathfrak{D}\Phi\right)_x+\Phi_{xxx}=0,
 \end{aligned}$$

(2)\ Supersymmetric MKdV equation \cite{Mathieu, Liu4, Ghosh2}
$$
 \begin{aligned}
 &\phi_{t}-3\phi\mathfrak{D}(\phi_x)\mathfrak{D}\phi-3(\mathfrak{D}\phi)^2\phi_x+\phi_{xxx}=0,
 \end{aligned}$$

(3) \ Supersymmetric It's  equation  \cite{Liu5}
$$\begin{aligned}
&\mathfrak{D}_tF_t+6(F_x(\mathfrak{D}_tF))_x+\mathfrak{D}_tF_{xxx}=0,
\end{aligned}$$

(4) \ Supersymmetric two-boson equation \cite{Brunelli, Zhang2}
$$
 \begin{aligned}
 &\phi_{1,t}=\mathfrak{D}((\mathfrak{D}\phi_1)^2)+2\phi_{2,x}-\phi_{1,xx},\\
 &\phi_{2,t}=2((\mathfrak{D}\phi_1)\phi_2)_x+\phi_{2,xx}.
 \end{aligned}$$

    The system (3.10) indicates that constructing multi-periodic wave solutions depends on
 the solvability of the system.  We consider the number of constraint  equations and some unknown parameters.
Obviously, the number of constraint  equations of the type (3.10) is
$2^N$. On the other hand we have parameters $\tau_{ii}, \tau_{ij},
\alpha_i, \omega_i, \phi_0, c$, whose total number is
$\frac{1}{2}N(N+1)+3N+2$. Among them, $2N$ parameters $\tau_{ii},
\omega_i$ are taken to be the given parameters related to the
amplitudes and wave numbers of $N$-periodic waves. Hence, the number
of the unknown parameters is $\frac{1}{2}N(N+1)+N+2$. while
$\frac{1}{2}N(N+1)$ parameters $\tau_{ij}$ implicitly appear in
series form, which is general can not to be solved explicit. Hence,
the number of the explicit unknown parameters is only $N+2$. The
number of equations is larger than the unknown parameters in the
case when $N> 2$.   This fact means that if equation (3.10) is
satisfied by the unknowns, we have at least $N$-periodic wave
solutions ($N\leq 2$). It is still possible to construct
multi-periodic wave solutions for $N\geq 3$ by adding the number of
parameters (for example, using constant solution and integration
constant) or decreasing the number of equations (for example, using
odd and even properties of considered equations). In this paper, we
consider one-periodic wave solution of the equation (1.2), which
belongs to the cases when $N=1$ and $N=2$  in the Riemann theta
function (3.1). There are still certain  difficulties in the
calculation for the case $N\geq 3$.

 We expect the proposed method to be extended to
$\mathcal{N}=1$ supersymmetric sine-Gordon equation and
$\mathcal{N}=1$ supersymmetric KP equation, as well as  other
discrete supersymmetric equations (like supersymmetric Toda
lattice). For the $\mathcal{N}=2$ supersymmetric equations with
ordinary variables $x,t$ and Grassmann variables
$\theta_1,\theta_2$, their corresponding superspace is
$\mathbb{R}_\Lambda^{2,2}=\Lambda_0^2\times\Lambda_1^2$. In this
case, the Grassmann algebra $G_2(\sigma_1, \sigma_2)$ whose
dimension is four. The $\tau$ matrix will be dependent on  the odd
parameters $\sigma_1, \sigma_2$. In superspace
$\mathbb{R}_\Lambda^{2,2}$, the super bilinear forms of
$\mathcal{N}=2$ supersymmetric equations, their quasi-periodic
solutions and asymptotic properties remain open.
  We intend to return to these question in some future publications.\\[12pt]
%%%%%%%%%%%%%%%%%%%%%%%%%%%%%%%%%%%%%%%%%%%%%%%%%%%%%%%%%%%%%%%%%%%%%%%%%
{\bf\large  Acknowledgment}

  The work  described in this paper was supported by grants from the National Science Foundation of China (No.10971031),
Shanghai Shuguang Tracking Project (No.08GG01) and Innovation
Program of Shanghai Municipal Education Commission (No.10ZZ131).

\end{document}